\documentclass[review]{elsarticle}

\usepackage{lineno,hyperref}
\modulolinenumbers[5]

\usepackage{graphicx}
\usepackage{deluxetable}
\usepackage{amsmath}

\journal{Icarus}

\begin{document}

\begin{frontmatter}

\title{The Concentric Maclaurin Spheroid method with tides and a rotational enhancement
of Saturn's tidal response}

\author[mymainaddress]{Sean M. Wahl\corref{mycorrespondingauthor} }
\cortext[mycorrespondingauthor]{Corresponding author}
\ead{swahl@berkeley.edu}

\author[mysecondaryaddress]{William B. Hubbard}

\author[mymainaddress,third]{Burkhard Militzer}

\address[mymainaddress]{Department of Earth and Planetary Science, The University of
California, Berkeley, CA, 94720-4767, USA} 
\address[mysecondaryaddress]{Lunar and Planetary Laboratory, The University of
Arizona, Tucson, AZ 85721-0092, USA}
\address[third]{Department of Astronomy, The University of California, Berkeley, CA,
94720-4767, USA}

\begin{abstract}
    We extend to three dimensions the Concentric Maclaurin Spheroid method for
    obtaining the self-consistent shape and gravitational field of a rotating liquid
    planet, to include a tidal potential from a satellite. We exhibit, for the first
    time, the important effect of the planetary rotation rate on tidal response of
    gas giants. Simulations of planets with fast rotation rates like those of Jupiter
    and Saturn, exhibit significant changes in calculated tidal love numbers $k_{nm}$
    when compared with non-rotating bodies. A test model of Saturn fitted to observed
    zonal gravitational multipole harmonics yields $k_2=0.413$, consistent with a
    recent observational determination from \textit{Cassini} astrometry data
    \citep{lainey2016}. The calculated love number is robust under reasonable
    assumptions of interior rotation rate, satellite parameters, and details of
    Saturn's interior structure. The method is benchmarked against several published
    test cases.
\end{abstract}


\end{frontmatter}

\linenumbers

\newcommand{\sphint}{\int_{-1}^1 d\mu' \int_{0}^{2\pi} d\phi' \int_{r' > r} dr'}
\newcommand{\sphshort}{\int_{\tau} d\tau}
\newcommand{\muint}{\int_{-1}^1 d\mu'}
\newcommand{\phiint}{\int_{0}^{2\pi} d\phi'}
\newcommand{\xiint}{\int_{b/a}^{\xi} d\xi'}

\section{Introduction} \label{intro}

The gas giants Jupiter and Saturn rotate so rapidly that adequate treatment of the
non-spherical part of their gravitational potential requires either a very high-order
perturbative, or better, an entirely non-perturbative approach \citep{hubbard2012,
hubbard2013, hubbard2014, wisdom1996,wisdom2016}. Here we present an extension of the
Concentric Maclaurin Spheroid (CMS) method of \citet{hubbard2012, hubbard2013} to three
dimensions to include the tidal perturbation from a satellite. This allows for
high-precision simulations of static tidal response, consistent with the
planet's shape and interior mass distribution. The presence of a large rotational
bulge produces an observable effect on the tidal response of giant planets. This
effect, which has not been previously revealed by linear tidal-response theories
applied to spherical-equivalent interior models, has implications for the observed
tidal responses of Jupiter and Saturn.

The \textit{Juno} spacecraft is expected to measure the strength of Jupiter's
gravitational field to an unprecedented precision ($\sim$ one part in $10^9$)
\citep{kaspi2010}, potentially revealing a weak signal from the planet's interior
dynamics. Also present in Jupiter's gravitational field will be tesseral-harmonic
terms produced by tides raised by the planet's large satellites.  In fact, close to
the planet, the gravitational signal from Jupiter's tides has a similar magnitude to
the predicted signal from models of deep internal dynamics
\citep{cao2015,kaspi2010,kaspi2013}. An accurate prediction of the planet's
hydrostatic tidal response will, therefore, be essential for interpreting the
high-precision measurements provided by the \textit{Juno} gravity science experiment.

Although the \textit{Cassini} Saturn orbiter was not designed for direct measurement
of high-order components of Saturn's gravitational field, it has already provided
gravitational information relevant to the planet's interior structure.
\citet{lainey2016} used an astrometry dataset of the orbits of Saturn's co-orbital
satellites to make the first determination of the planet's $k_2$ love number. Their
observed $k_2$ was significantly larger than the theoretical prediction of
\citet{gavrilov1977}. A mismatch between an observed $k_2$ and the value predicted
for a Saturn model fitted to the planet's low-degree zonal harmonics $J_2$ and $J_4$
would raise questions about the adequacy of the hydrostatic (non-dynamic) theory of
tides.  

In this paper we present theoretical results for simplified Saturn interior models
matching the planet's observed low-degree zonal harmonics.  When these models are
analyzed with the full 3-d CMS theory including rotation and tides, we predict a
gravitational response in line with the observed $k_2$ value of \citet{lainey2016},
suggesting that the observation can be completely understood in terms of a static
tidal response.  A similar test will be possible for Jupiter once its $k_2$ has been
measured by the \textit{Juno} spacecraft.

There is extensive literature on the problem of the shape and gravitational potential
of a liquid planet in hydrostatic equilibrium, responding to its own rotation and to
an external gravitational potential from a satellite; see, e.g., a century-old
discussion in \citet{jeans1919}.  Many classical geophysical investigations use a
perturbation approach, obtaining the planet's linear and higher-order response to
small deviations of the potential from spherical symmetry. A good discussion of the
application of perturbation theory to rotational response, the so-called theory of
figures, is found in \citet{zharkov1978}, while a pioneering calculation of the tidal
response of giant planets is presented by \citet{gavrilov1977}.

\citet{hubbard2012} introduced an iterative numerical method, based on the theory of
figures, for calculating the self-consistent shape and gravitational field of a
constant density, rotating fluid body to high precision. In this method, integrals
over the mass distribution are solved using Gaussian quadrature to obtain the
gravitational multipole moments. This method was extended to non-constant density
profiles by \citet{hubbard2013}, by approximating the barotropic pressure-density
relationship with multiple concentric Maclaurin (i.e., constant-density) spheroids.
This approach (called the CMS method) mitigates problems with cancellation of terms
that arise in a purely numerical solution to the general equation of hydrostatic
equilibrium, and has a typical relative precision of $\sim 10^{-12}$. The CMS method
has been benchmarked against analytical results for simple models \citep{hubbard2014}
and against an independent, non-perturbative numerical method
\citep{wisdom1996,wisdom2016}. 

The theory of \citet{gavrilov1977} begins with an interior model of Saturn fitted to
the values of $J_2$ and $J_4$ observed at that time.  This interior model tabulates
the mass density $\rho$ as a function of $s$, where $s$ is the mean radius of the
constant-density surface.  Tidal perturbation theory is then applied to this
spherical-equivalent Saturn.  The \citet{gavrilov1977} approach is sufficient for an
initial estimate of the tidally-induced terms in the external potential, but it
neglects terms which are of the order of the product of the tidal perturbation and
the rotational perturbation.  Here we demonstrate that, for a rapidly-rotating giant
planet, the latter terms make a significant contribution to the love numbers
$k_{nm}$, as well as (unobservably small) tidal contributions to the gravitational
moments $J_n$.

\citet{folonier2015} presented a method for approximating the love numbers of a
non-homogeneous body using Clairaut theory for the equilibrium ellipsoidal figures.
This results in an expression for the love number $k_2$ for a body composed of
concentric ellipsoids, parameterized by their flattening parameters. In the case of
the constant density Maclaurin spheroid, there is a well-known result that the
equipotential surface is an ellipsoid. However, in bodies with more complicated
density distributions, the equipotential surfaces will have a more general spheroidal
shape.  Because of the small magnitude of tidal perturbations, the method of
\citet{folonier2015} works in the limit of slow rotation despite this limitation.
However, the method does not account for the coupled effect of tides and rotation, and
does not predict love numbers of order higher than $k_2$.  Within these constraints,
we show below that our extended CMS method yields results that are in excellent
agreement with results from \citet{folonier2015}.

Although our theory is quite general and can be used to calculate a rotating planet's
tidal response to multiple satellites located at arbitrary latitudes, longitudes, and
radial distances, for application to Jupiter and Saturn it suffices to consider the
effect of a single perturbing satellite sitting on an orbital plane at zero
inclination to the planet's equator.  Since tidal distortions are always very small
compared with rotational distortion, and Jupiter's Galilean satellites, as well many
of Saturn's larger satellites, are on orbits with low inclination, the tidal response
to multiple satellites can be obtained by a linear superposition of the perturbation
from each body.  Extension of our theory to a system with a large satellite on an
inclined orbit, such as Neptune-Triton, would be straightforward, but is not
considered here.

\section{Concentric Maclaurin Spheroid method with tides} \label{methods}

\subsection{Model parameters} \label{parameters}

In the co-rotating frame of the planet in hydrostatic equilibrium,  the pressure $P$,
the mass density $\rho$ and the total effective potential $U$ are related by
\begin{equation} \begin{aligned} \nabla P = \rho \nabla U. \end{aligned}
    \label{eq:hydrostatic} \end{equation}
The total effective potential can be separated into three components,
\begin{equation} \begin{aligned} U = V + Q + W, \end{aligned}
    \label{eq:potential_components} \end{equation}
where $V$ is the gravitational potential arising from the mass distribution within
the planet, $Q$ is the centrifugal potential corresponding to a rotation frequency
$\omega$, and $W$ is the tidal potential arising from a satellite with mass $m_{\rm
s}$ at planet-centered coordinates $(R,\mu_s,\phi_s)$, where $R$ is the satellite's
orbital distance from the origin, $\mu_s=\cos \theta$, where $\theta$ is the satellite's
planet-centered colatitude and $\phi_s$ is the planet-centered longitude.  For the
purposes of this investigation, we always place the satellite at angular coordinates
$\mu_s=0$ and $\phi_s=0$.  The relative magnitudes of $V$, $Q$, and $W$ can be
described in terms of two non-dimensional numbers:
\begin{equation} q_{\rm rot} = \frac{\omega^2 a^3}{GM} \label{eq:qrot} \end{equation}
for the rotational perturbation and
\begin{equation} q_{\rm tid} = -\frac{3m_{\rm s}a^3}{MR^3} \label{eq:qtid}
\end{equation}
for the tidal perturbation, where $G$ is the universal gravitational constant, and
$M$ and $a$ are the mass and equatorial radius of the planet. The planet-satellite
system is described by these two small parameters along with a third parameter, the
ratio $a/R$. 

Since CMS theory is nonperturbative, in principle our results are valid to all powers
of these small parameters and their products (until we reach the computer's numerical
precision limit). For the giant-planet tidal problems that we consider here, terms of
second and higher order in $q_{\rm tid}$ are always negligible, but terms linear in
$q_{\rm tid}$ and multiplied by various powers of $q_{\rm rot}$ and $a/R$ contribute
above the numerical noise level.  It is, in fact, terms of order $q_{\rm tid} \cdot
q_{\rm rot}$ that contribute most importantly to the new results of this paper.

\begin{figure}[h!]  
  \centering
    \includegraphics[width=1.0\textwidth]{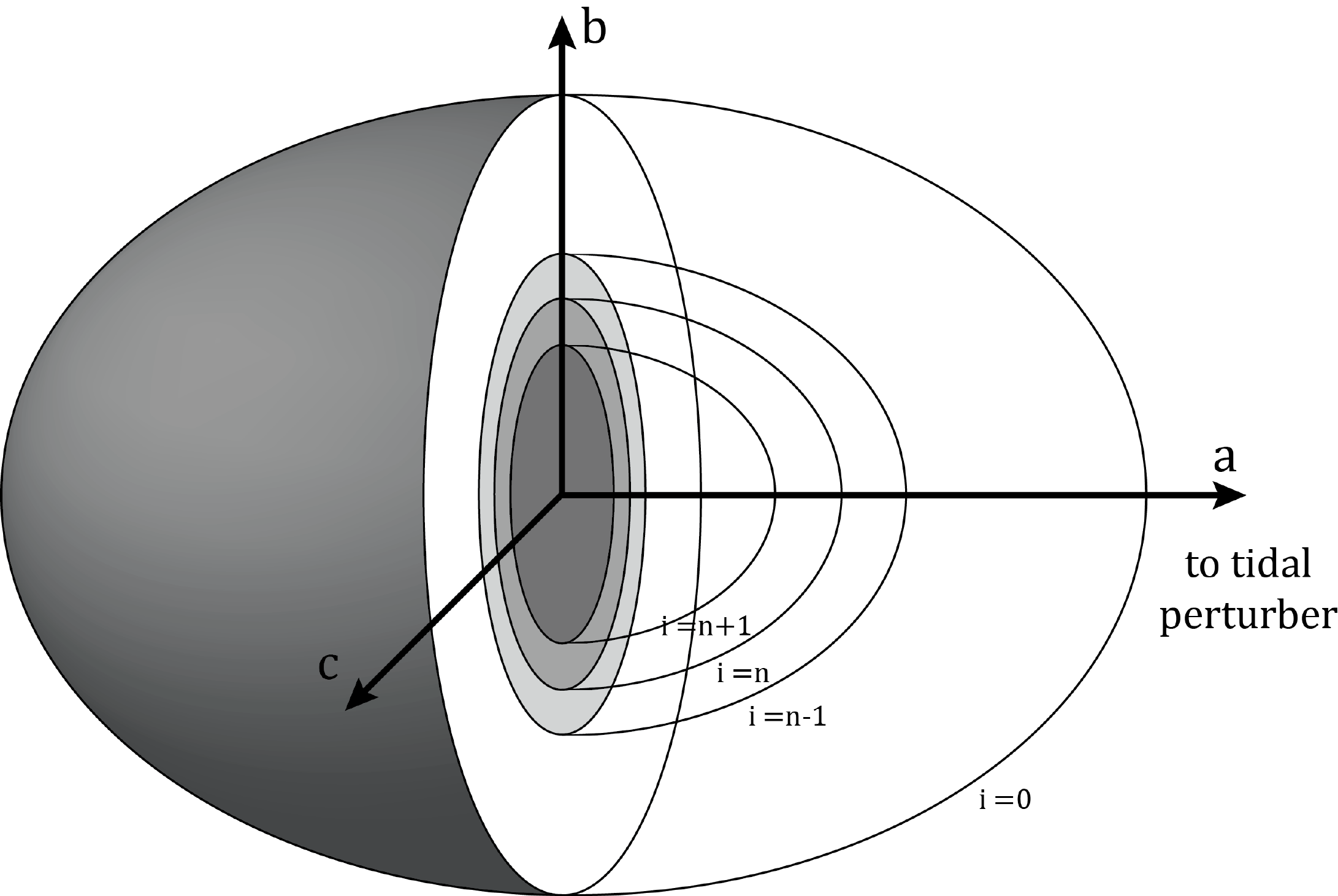}
\caption{ Conceptual diagram of a Concentric Maclaurin Spheroid model with a tidal
perturbation from a satellite.}
\label{fig:spheroid}
\end{figure}

We introduce dimensionless planetary units of pressure $P_{\rm pu}$, density
$\rho_{\rm pu}$, and total potential $U_{\rm pu}$, such that
\begin{equation} \begin{aligned} P \equiv& \frac{GM^2}{a^4} P_{\rm pu} \\ \rho
        \equiv& \frac{M}{a^3} \rho_{\rm pu} \\ U \equiv& \frac{GM}{a} U_{\rm pu}. \\
    \end{aligned} \label{eq:planetary_units} \end{equation}
The CMS method considers a model planet composed of $N$ nested spheroids of constant
density as depicted in Figure \ref{fig:spheroid}. We label these spheroids with index
$i=0,1,2,\dots,N-1$, with $i = 0$ corresponding to the outermost spheroid and $i=N-1$
corresponding to the innermost spheroid. Each spheroid is constrained to have a point
at radial distance $a_i$ from the planet's center of mass, such that each of these
fixed points has the same angular coordinates as the sub-satellite point
$(\mu=0,\phi=0)$.  Accordingly, the $a_0$ of the outermost spheroid corresponds to
its the largest principal axis, if the perturbing satellite is in the equatorial
plane.

When $q_{\rm tid}=0$, the potential is axially symmetric and the problem can be
solved in two spatial dimensions. However, when both $q_{\rm tid}$ and $q_{\rm rot}$
are nonzero, the symmetry is broken, meaning that each spheroid has a fully triaxial
figure with the surface described by
\begin{equation} \zeta_i \equiv r_i(\mu,\phi) / a_i, \label{eq:shape} \end{equation}
such that $\zeta_0$ represents the shape of the outer surface.

Taking advantage of the principle of superposition for a linear relationship between
the potential $V$ and the mass density $\rho$, the total $V$ is given by the sum of
the potential arising from each individual spheroid \citep{hubbard2013}. This allows us to
approximate any monotonically increasing density profile, with the density of the
$i$th spheroid represented by the density jump
\begin{equation} \delta \rho_i = \begin{cases} \rho_i - \rho_{i-1}, & i>0 \\ \rho_0,
        & i=0.  \end{cases} \label{eq:density_increment} \end{equation}
This parameterization of density has the added benefit of naturally handling
discontinuities in $\rho$, as would be expected for a giant planet with a dense
central core.

\subsection{Calculation of gravitational potential} \label{method_derivation}

The general expansion of $V$ in spherical coordinates ${\bf r} = (r,\mu =
\cos{\theta},\phi)$ is \begin{equation} \begin{aligned} V&(r,\mu,\phi) = \\
        &\frac{GM}{r} \left[ \sum_{n=0}^{\infty} P_n(\mu) \sphshort
            \rho(r')P_n(\mu')\left(\frac{r'}{r}\right)^k \right. \\ & +
            \sum_{n=0}^{\infty} \sum_{m=1}^{n} P^m_n(\mu)\cos(m\phi) \sphshort
            \frac{2(n-m)!}{(n+m)!} \rho(r')P_n^m(\mu')\cos(m\phi')
            \left(\frac{r'}{r}\right)^k \\ & \left.  + \sum_{n=0}^{\infty}
            \sum_{m=1}^{n} P^m_n(\mu)\sin(m\phi) \sphshort \frac{2(n-m)!}{(n+m)!}
            \rho(r')P_n^m(\mu')\sin(m\phi') \left(\frac{r'}{r}\right)^k \right]
        \end{aligned} \label{eq:general_expansion} \end{equation}
\citep{zharkov1978}, where $P_n$ and $P_n^m$ are the Legendre and associated Legendre
polynomials, 
\begin{equation*} d\tau = r'^2 \sin(\theta')d\theta' d\phi' = r'^2 d\mu' d\phi',
\end{equation*}
and the origin, ${\bf r}=(0,0,0)$, is the center of mass of the planet.  The
potential at a general point within the planet has a contribution from mass both
interior and exterior to that point, for which the exponent $k$ in Eqn.
\eqref{eq:general_expansion} is different:
\begin{equation*} k = \begin{cases} n, & r' < r \\ -(n+1), & r' > r .  \end{cases}
\end{equation*}

The centrifugal potential $Q$ depends only on $r$ and $\mu$
\begin{equation} \begin{aligned} Q(r,\mu) = \frac{1}{3}r^2\omega^2\left[1-P_2(\mu)
        \right].  \end{aligned} \label{eq:centrifugal} \end{equation}
The tidal potential $W$ for a satellite at position ${\bf R}=(R,\mu_s,\phi_s)$ is
\begin{equation} \begin{aligned} W({\bf r}) = \frac{Gm_{\rm s}}{\left| {\bf R} - {\bf
        r} \right|}.  \end{aligned} \label{eq:tidal} \end{equation}
The general expansion of $W$ around the center of mass of the planet is obtained by
using the summation theorem for spherical harmonics \citep{gavrilov1977}  
\begin{equation} \begin{aligned} W(r,\mu,\phi) =& \frac{Gm_{\rm s}}{R}
        \sum_{n=0}^\infty \left[ P_n(\mu) P_n(\mu_s) \phantom{\frac{1}{1}}\right. \\
        & \left. + 2 \sum_{m=1}^{n} \frac{2(n-m)!}{(n+m)!} \cos(m\phi - m\phi_s)
    P_n^m(\mu) P_n^m(\mu_s) \right].  \end{aligned} \label{eq:tidal_general}
    \end{equation}

Following \citet{hubbard2013}, we derive non-dimensional quantities in terms of the
planet mass $M$ and maximum radius $a=a_0$. For each spheroid, we define a
dimensionless radius of each spheroid
\begin{equation} \begin{aligned} \lambda_i \equiv& a_i/a \\
    \end{aligned} \label{eq:lambda} \end{equation}
and dimensionless density increment, based on the mean density of the planet
\begin{equation} \begin{aligned} \bar{\rho} =& \frac{3M}{a^3}\frac{1}{\muint \phiint
        \zeta_0^3} \\ \delta_i \equiv& \frac{\delta\rho_i}{\bar{\rho}}.
        \label{eq:delta} \end{aligned} \end{equation}
The model planet's mass is then given by the integral expression
\begin{equation} M = \frac{1}{3} \sum_{j=0}^{N-1} \delta_j \lambda_j^3 \muint \phiint
    \zeta_j^3.  \label{eq:mass} \end{equation}
The contribution to the potential is expanded in terms of interior and external zonal
harmonics $J_{i,n}$ and $J_{i,n}'$. For the tidal problem, we must also consider the
analogous $C_{i,nm}$, $C'_{i,nm}$, $S_{i,nm}$ and $S'_{i,nm}$. These contribute
linearly to the total moment evaluated exterior to the planet's surface; for
instance,
\begin{equation} J_2 = \sum_{i=0,N-1} J_{i,2}.  \label{eq:total_J2} \end{equation}
%
\newcommand{\til}{\widetilde}
The layer-specific harmonics are then normalized by radius as 
\begin{equation} \begin{aligned}[c] \til{J}_{i,n} \equiv&
        \frac{J_{i,n}}{\lambda_i^n}, & \til{J}'_{i,n} \equiv&
        J'_{i,n}\lambda_i^{(n+1)} \\ \til{S}_{i,nm} \equiv&
        \frac{S_{i,nm}}{\lambda_i^n}, & \til{S}'_{i,nm} \equiv&
        S'_{i,nm}\lambda_i^{(n+1)} \\ \til{C}_{i,nm} \equiv&
        \frac{C_{i,nm}}{\lambda_i^n}, & \til{C}'_{i,nm} \equiv&
        C'_{i,nm}\lambda_i^{(n+1)}.  \end{aligned} \label{eq:normalization}
\end{equation}
Following the derivation in \citet{hubbard2013} and generalizing the expressions for
full three dimensional volume integrals, we find the normalized interior harmonics
\begin{equation} \begin{aligned} \til{J}_{i,n} &=& -\frac{3}{n+3} \frac{ \delta_i
            \lambda_i^3 \muint P_n(\mu') \phiint \zeta_i^{(n+3)} } { \sum_{j=0}^{N-1}
        \delta_j \lambda_j^3 \muint \phiint \zeta_j^3} \\ \til{C}_{nm} &=&
        \frac{6(n-m)!}{(n+3)(n+m)!} \frac{ \delta_i \lambda_i^3 \muint P_n^m(\mu')
            \phiint \zeta_i^{(n+3)} \cos(m\phi') } { \sum_{j=0}^{N-1} \delta_j
        \lambda_j^3 \muint \phiint \zeta_j^3} \\ \til{S}_{nm} &=&
        \frac{6(n-m)!}{(n+3)(n+m)!} \frac{ \delta_i \lambda_i^3 \muint P_n^m(\mu')
            \phiint \zeta_i^{(n+3)} \sin(m\phi') } { \sum_{j=0}^{N-1} \delta_j
        \lambda_j^3 \muint \phiint \zeta_j^3}, \\ \end{aligned} \label{eq:harmonics}
\end{equation}
and the exterior harmonics
\begin{equation} \begin{aligned} \til{J}'_{i,n} &=& -\frac{3}{2-n} \frac{ \delta_i
            \lambda_i^3 \muint P_n(\mu') \phiint \zeta_i^{(-n+2)} } {
                \sum_{j=0}^{N-1} \delta_j \lambda_j^3 \muint \phiint \zeta_j^3} \\
                \til{C}'_{nm} &=&  \frac{6(n-m)!}{(2-n)(n+m)!} \frac{ \delta_i
                    \lambda_i^3 \muint P_n^m(\mu') \phiint \zeta_i^{(-n+2)}
                \cos(m\phi') } { \sum_{j=0}^{N-1} \delta_j \lambda_j^3 \muint \phiint
            \zeta_j^3} \\ \til{S}'_{nm} &=& \frac{6(n-m)!}{(2-n)(n+m)!} \frac{
                \delta_i \lambda_i^3 \muint P_n^m(\mu') \phiint  \zeta_i^{(-n+2)}
            \sin(m\phi') } { \sum_{j=0}^{N-1} \delta_j \lambda_j^3 \muint \phiint
        \zeta_j^3} \\ \end{aligned} \label{eq:harmonics_prime} \end{equation}
with a special case for $n=2$
\begin{equation} \begin{aligned} \til{J}'_{i,n} &=& -3 \frac{ \delta_i \lambda_i^3
        \muint P_n(\mu') \phiint \log(\zeta_i) } { \sum_{j=0}^{N-1} \delta_j
    \lambda_j^3 \muint \phiint \zeta_j^3} \\ \til{C}'_{nm} &=&
    \frac{6(n-m)!}{(n+m)!} \frac{ \delta_i \lambda_i^3 \muint P_n^m(\mu') \phiint
    \log(\zeta_i) \cos(m\phi') } { \sum_{j=0}^{N-1} \delta_j \lambda_j^3 \muint
\phiint \zeta_j^3} \\ \til{S}'_{nm} &=& \frac{6(n-m)!}{(n+m)!} \frac{ \delta_i
\lambda_i^3 \muint P_n^m(\mu') \phiint  \log(\zeta_i) \sin(m\phi') } {
    \sum_{j=0}^{N-1} \delta_j \lambda_j^3 \muint \phiint \zeta_j^3} \\ \end{aligned}
\label{eq:harmonics_prime_n=2} \end{equation}
and
\begin{equation} \begin{aligned} J''_{i,0} =& \frac{2\pi\delta_i a_0^3}{3 M}.
    \end{aligned} \end{equation}

The shape of the surface of the planet is defined by the equipotential relationship
\begin{equation} U(\zeta,\mu,\phi,\mu_s,\phi_s) -  U(1,0,0,\mu_s,\phi_s) = 0,
    \label{eq:equipotential_surface} \end{equation}
where the potential in planetary units at an arbitrary point on the planet's surface
\begin{equation} \begin{aligned} U(\zeta,\mu,\phi,\mu_s,\phi_s) =& \frac{1}{\zeta_0}
        \left[ 1 - \sum^{N-1}_{i=0} \sum^\infty_{n=1} \lambda_i^n \zeta_0^{-n}
            \left\{ P_{n}(\mu) \til{J}_{i,n} \phantom{\sum_{1}^{n}}\right. \right. \\
            & \left. - \sum_{m=1}^{n} P_n^m(\mu) \left( \til{C}_{i,nm}\cos(m\phi) +
            \til{S}_{i,nm}\sin(m\phi) \right) \right\}  \\ & + \frac{1}{3} q_{\rm
            rot} \zeta_0^3 [1 - P_2(\mu)] \\ & -\frac{1}{3} \zeta_0^{3} q_{\rm tid}
            \sum_{n=2}^\infty \left(\frac{a}{R}\right)^{(n-2)} \zeta_0^{(n-2)}
            \left\{ P_n(\mu) P_n(\mu_s) \phantom{\sum_1^n}\right. \\ & \left. \left.
        + 2 \sum_{m=1}^{n} \frac{(n-m)!}{(n+m)!} \cos(m\phi - m\phi_s) P_n^m(\mu)
    P_n^m(\mu_s) \right\} \right]\\ \end{aligned} \label{eq:surface_potential}
    \end{equation}
matches the reference potential at the sub-satellite point
\begin{equation} \begin{aligned} U(1,0,0,\mu_s,\phi_s) =& 1 - \sum^{N-1}_{i=0}
        \sum^\infty_{n=1} \lambda_i^n   \left\{ P_{n}(0) \til{J}_{i,n} -
        \sum_{m=1}^{n} P_n^m(0) \til{C}_{i,nm} \right\}  \\ &+ \frac{1}{2} q_{\rm
        rot}  -\frac{1}{3} q_{\rm tid} \sum_{n=2}^\infty
        \left(\frac{a}{R}\right)^{(n-2)} \left\{ P_n(0) P_n(\mu_s) \phantom{\sum_a^b}
        \right. \\ & \left. + 2 \sum_{m=1}^{n} \frac{(n-m)!}{(n+m)!} \cos(- m\phi_s)
    P_n^m(0) P_n^m(\mu_s) \right\}. \\ \end{aligned} \label{eq:surface_subsatellite}
    \end{equation}
Similarly, the shapes of the interior spheroids are found by solving
\begin{equation} U_j(\zeta,\mu,\phi,\mu_s,\phi_s) -  U_j(1,0,0,\mu_s,\phi_s) = 0,
    \label{eq:equipotential_interior} \end{equation}
where
\begin{equation} \label{eq:Uj} \begin{aligned} U_j(\zeta_j,\mu,\phi,\mu_s,\phi_s) =&
        -\frac{1}{\zeta_j \lambda_j} \left[ \sum^{N-1}_{i=j} \sum^\infty_{n=0} \left(
            \frac{\lambda_i}{\lambda_j} \right)^n \zeta_j^{-n} \left\{ P_{n}(\mu)
            \til{J}_{i,n} \phantom{\sum_a^b} \right. \right. \\ & \left. -
            \sum_{m=1}^{n} P_n^m(\mu) \left( \til{C}_{i,nm}\cos(m\phi) +
            \til{S}_{i,nm}\sin(m\phi) \right) \right\}  \\ &+ \sum^{j-1}_{i=0}
            \sum^\infty_{n=0} \left( \frac{\lambda_j}{\lambda_i} \right)^{n+1}
            \zeta_j^{n+1} \left\{ \til{J}'_{i,n} P_n(\mu) \phantom{ \sum_1^m} \right.
            \\ & \left. - \sum_{m=1}^{n} P_n^m(\mu) \left( \til{C}'_{i,nm}\cos(m\phi)
        + \til{S}'_{i,nm}\sin(m\phi) \right) \right\} \\ &+ \left. \sum^{j-1}_{i=0}
    J''_{i,0} \lambda_j^3 \zeta_j^3 \right] + \frac{1}{3} q_{\rm rot} \lambda_j^2
    \zeta_j^2 [1 - P_2(\mu)] \\ &-\frac{1}{3} \lambda_j^2 \zeta_j^{2} q_{\rm tid}
    \sum_{n=2}^\infty \left(\frac{a\lambda_j}{R}\right)^{(n-2)} \zeta_j^{(n-2)}
    \left\{ P_n(\mu) P_n(\mu_s) \phantom{\sum_a^b} \right. \\ & \left. + 2
    \sum_{m=1}^{n} \frac{(n-m)!}{(n+m)!} \cos(m\phi - m\phi_s) P_n^m(\mu)
P_n^m(\mu_s) \right\} \end{aligned} \end{equation}
and
\begin{equation} \begin{aligned} U_{j}(1,0,0,\mu_s,\phi_s) =& -
        \frac{1}{\lambda_j}\left[ \sum^{N-1}_{i=j} \sum^\infty_{n=0} \left(
            \frac{\lambda_i}{\lambda_j} \right)^n   \left\{ P_{n}(0) \til{J}_{i,n} -
            \sum_{m=1}^{n} P_n^m(0) \til{C}_{i,nm}\right\} \right.\\ & +
        \sum^{j-1}_{i=0} \sum^\infty_{n=0} \left( \frac{\lambda_j}{\lambda_i}
    \right)^{n+1} \left\{ \til{J}'_{i,n} P_n(0) - \sum_{m=1}^{n} P_n^m(0)
\til{C}'_{i,nm} \right\} \\ & \left. + \sum^{j-1}_{i=0} J''_{i,0} \lambda_j^3 \right]
+\frac{1}{2} q_{\rm rot} \lambda_j^2 \\ & - \frac{1}{3} \lambda_j^2 q_{\rm tid}
\sum_{n=2}^\infty \left(\frac{a\lambda_j}{R}\right)^{(n-2)} \left\{  P_n(0)
    P_n(\mu_s) \phantom{\sum_1^1} \right. \\ & \left. + 2 \sum_{m=1}^{n}
    \frac{(n-m)!}{(n+m)!} \cos(- m\phi_s) P_n^m(0) P_n^m(\mu_s) \right\}. \\
\end{aligned} \label{eq:interior_subsatellite} \end{equation}
From Eqn. \eqref{eq:interior_subsatellite}, we also find the potential at the center
of the planet
\begin{equation} \begin{aligned} U_{\rm center} = -\sum_{i=0}^{N-1} \sum_{n=0}^\infty
        \lambda_i \left\{ \til{J}'_{i,n} - \sum_{m=1}^{n} \til{C}'_{i,nm} \right\}.
    \end{aligned} \label{eq:central_potential} \end{equation}

Taking the limit of Eqn. \eqref{eq:central_potential} as the radius goes to zero
yields
\begin{equation} \begin{aligned} U_{\rm center} &=\lim_{\zeta_j\to 0} U_j(\zeta_j)\\
        &= - \sum_{i=0}^{N-1} \frac{J'_{i,n=0}}{ \lambda_i}, \end{aligned}
\end{equation} correcting a typographical error in Eqn. 49 of \citet{hubbard2013}.
In solving equations \eqref{eq:equipotential_surface} and
\eqref{eq:equipotential_interior}, we also require their analytical derivatives
\begin{equation} \begin{aligned} \frac{d \left[U(\zeta,\mu,\phi,\mu_s,\phi_s) -
        U(1,0,0,\mu_s,\phi_s) \right] }{d\zeta} =& \frac{dU(\zeta,\mu,\phi)}{d\zeta}
        \\ \frac{d \left[ U_j(\zeta_j,\mu,\phi,\mu_s,\phi_s) -
        U_{j}(1,0,0,\mu_s,\phi_s) \right]}{d\zeta_j} =& \frac{d
        U_j(\zeta_j,\mu,\phi)}{d\zeta_j}.  \label{eq:equipotential_derivative}
    \end{aligned} \end{equation}

\subsection{Gaussian quadrature} \label{quadrature}

The preceding expressions give the gravitational potential and equipotential shapes,
as a function of $q_{\rm rot}$ and $q_{\rm tid}$, within a layered planet with $N$
concentric spheroids. In the limit of $N\to \infty$, the solution would apply to an
arbitrary monotonically increasing barotropic relation, $\rho(P)$. 

For practical applications, we need to find the potential as a multipole expansion up
to a maximum degree $n_{\max}$. For the results presented here, we use $n_{\max}=30$.
The angular integrals in equations
\eqref{eq:harmonics}~--~\eqref{eq:harmonics_prime_n=2} can be evaluated using
Gaussian quadratures on a two dimensional grid. Here we use Legendre-Gauss
integration to integrate polar angles over $L_{1}=48$ quadrature points
$\mu_{\alpha}=\cos(\theta_{\alpha})$, $\alpha = 1,2,\dots L_{1}$, with the
corresponding weights $\omega_{\alpha}$, $\alpha = 1,2,\dots L_{1}$ over the interval
$0<\mu<1$. At any point in the calculation, we must keep track of radius values for
each layer on a 2D grid of quadrature points $\zeta_{i\alpha\beta}$. For efficiency,
we precalculate the values of all of the Legendre and associated Legendre polynomials
at each polar quadrature point, $P_n(\mu_\alpha)$ and $P_n^m(\mu_\alpha)$.

For the azimuthal angle, we encounter integrals of the form
\begin{equation} \begin{aligned} I_{c,m} \equiv \int^{2\pi}_0 f(\phi)
        \cos(m\phi)d\phi \\ I_{s,m} \equiv \int^{2\pi}_0 f(\phi) \sin(m\phi)d\phi
    \end{aligned} \label{eq:cheb_integral} \end{equation}
when calculating the tesseral harmonics. For these, we use Chebyshev-Gauss
integration with $L_{2}=96$ quadrature points $\eta_{\beta}=\cos(\phi_{\beta})$,
$\beta = 1,2,\dots L_{2}$, with the corresponding weights $\omega_{\beta}$, $\beta =
1,2,\dots L_{2}$ over the interval $0<\phi<2\pi$  
\begin{equation} \begin{aligned} d\eta = -\sin(\phi)d\phi \\ d\phi = -
        \frac{d\eta}{\sqrt{1-\eta^2}}.  \end{aligned} \end{equation}
Using the identity $(\sin\theta)^{m-k}=(1-\mu^2)^{\frac{m-k}{2}}$, the sinusoidal
functions can be expanded as
\begin{equation} \begin{aligned} \cos{m\phi} = \sum^m_{k=0} { m \choose k } \eta^k
        (1-\eta^2)^{\frac{m-k}{2}} \cos\left\{ \frac{\pi}{2}(m-k) \right\} \\
        \sin{m\phi} = \sum^m_{k=0} { m \choose k } \eta^k (1-\eta^2)^{\frac{m-k}{2}}
        \sin\left\{ \frac{\pi}{2}(m-k) \right\}.  \end{aligned} \end{equation}
Substituting these into Eqn. \eqref{eq:cheb_integral} and splitting the integral into
two intervals $0<\phi<\pi$ and $\pi<\phi<2\pi$ yields
\begin{equation} \begin{aligned} I_{c,m} =& \sum^m_{k=0}{ m \choose k } \cos \left[
        \frac{\pi}{2}(m-k) \right] \left\{ \int^{1}_{-1} \eta^k f(\cos^{-1}(-\eta))
        \left[ 1 - \eta^2 \right]^{\frac{m-k}{2}} d\eta \right. \\ &- \left.
        \int^{1}_{-1} \eta^k f(\cos^{-1}\eta) \left[ 1 - \eta^2
        \right]^{\frac{m-k}{2}} d\eta \right\} \\ =& \sum^m_{k=0}{ m \choose k } \cos
        \left[ \frac{\pi}{2}(m-k) \right] \\ &* \left\{ \pm \sum^{L_2}_{\beta =1}
        \omega_\beta \eta_{\beta}^k f(\pi - \cos^{-1}(\eta_{\beta})) \left[ 1 -
        \eta_\beta^2 \right]^{\frac{m-k}{2}} \right. \\ &- \left. \sum^{L_2}_{\beta
        =1} \omega_\beta \eta_\beta^k f(\cos^{-1}\eta_\beta) \left[ 1 - \eta_\beta^2
        \right]^{\frac{m-k}{2}} \right\}, \end{aligned} \end{equation}
where the sign of the second sum depends on the parity of $m$.  When calculating the
zonal harmonics, the integral $I_{c,m}(f(\mu_\alpha,\phi_\beta))$ reduces to the
axisymmetric solution with $m=0$. The zonal harmonics Eqn.  \eqref{eq:harmonics} can,
therefore, be calculated via the summation
\begin{equation} \begin{aligned} \til{J}_{i,n} \approx - \left( \frac{3}{n+3} \right)
        \left( \frac{ \delta_i \lambda_i^3 \sum^{L_1}_{\alpha =1} \omega_{\alpha}
        P_n(\mu_{\alpha}) I_{c,0}( \zeta_{i\alpha\beta}^{(n+3)})} { \sum_{j=0}^{N-1}
        \delta_j \lambda_j^3  \sum^{L_1}_{\alpha =1} \omega_{\alpha} I_{c,0}(
        \zeta_{j\alpha\beta}^{3} )} \right) \end{aligned} \label{eq:zonal_quad}
\end{equation}
and the tesseral harmonics likewise via
\begin{equation} \begin{aligned} \til{C}_{nm} &\approx&  \frac{6(n-m)!}{(n+3)(n+m)!}
        \left( \frac{ \delta_i \lambda_i^3 \sum^{L_1}_{\alpha =1} \omega_{\alpha}
        P^m_n(\mu_{\alpha})  I_{c,m}(\zeta_{i\alpha\beta}^{(n+3)}) } {
            \sum_{j=0}^{N-1} \delta_j \lambda_j^3 \sum^{L_1}_{\alpha =1}
            \omega_{\alpha} I_{c,0}(\zeta_{j\alpha\beta}^3)} \right).  \end{aligned}
        \label{eq:tesseral_quad} \end{equation}
There are analogous expressions for $I_{s,m}$ and $S_{nm}$, but these evaluate to
zero in all calculations presented here due to the symmetry of the model.

\subsection{Iterative procedure} \label{iterative}

We begin with initial estimates for the shape of each surface
$\zeta_{i\alpha\beta,0}$ and for the moments $\til{J}_{i,n}$, $\til{J}'_{i,n}$,
$\til{J}''_i$, $\til{C}_{i,nm}$, $\til{C}'_{i,nm}$, $\til{S}_{i,nm}$, and
$\til{S}'_{i,nm}$. For each iteration $t$ the level surfaces are then updated using a
single Newton-Raphson integration step. 
\begin{equation} \begin{aligned} \zeta_{i\alpha\beta,t+1} = \zeta_{i\alpha\beta,t} -
        \frac{ f( \zeta_{i\alpha\beta,t})}{ f'( \zeta_{i\alpha\beta,t}) }
    \end{aligned} \label{eq:newton} \end{equation}
where $f$ is the equipotential relation, Equations
\eqref{eq:equipotential_surface}~--~\eqref{eq:surface_subsatellite} for the outermost
surface and Equations
\eqref{eq:equipotential_interior}~--~\eqref{eq:interior_subsatellite} for interior
layers, and $f'$ is the first derivative of that function with respect to $\zeta$, Eqn.
\eqref{eq:equipotential_derivative}. The multipole moments are then calculated for
the updated $\zeta_{i\alpha\beta}$ via Equations
\eqref{eq:harmonics}~--~\eqref{eq:harmonics_prime_n=2}. These two steps are repeated
until all of the exterior moments, $J_{n}$, $C_{nm}$ and $S_{nm}$,
have converged such that the difference between successive iterations falls below a
specified tolerance. Starting with a naive guess for the initial state, a typical
calculation achieves a precision much higher than would be required for comparison
with \textit{Juno} measurements after about 40 iterations.

In simulations with a finite $q_{\rm rot}$ and $q_{\rm tid}$, we typically find an
initial converged equilibrium shape with a non-zero, first-order harmonic coefficient
$C_{11}$ of the order of $q_{\rm rot} \cdot q_{\rm tid}$ or smaller. This indicates
that the center of mass of the system is shifted slightly along the planet-satellite
axis from the origin of the initial coordinate system. To remove this term, we apply
a translation to the shape function of $\Delta x=-a\cdot C_{11}$ in the direction of
the satellite. This correction requires approximating the coordinates $(\mu',\phi')$
in the uncorrected frame that correspond to the quadrature points $\mu_\alpha$ and
$\phi_\beta$ in the corrected frame, so that the correct shape $\zeta$ is integrated
to find the moments in the corrected frame. For a value of $q_{\rm tid}$ similar to
the gas giants, this correction yields a body with $C_{11}$ on the order of the
specified tolerance. For systems with a much larger $q_{\rm tid}$ (of which there are
none in our planetary system), this second-order effect might affect the precision of
the calculation.  The residual effect is below the numerical noise level for the
Saturn models presented in this paper.

\subsection{Calculation of the barotrope} \label{barotrope}

We first calculate the density of each uniform layer; for the $j$th layer we have 
\begin{equation}
\begin{aligned}
    \rho_{j,{\rm pu}} = \frac{\sum_{i=0}^j\delta_i}{\sum_{k=0}^{N-1}
        \delta_k \lambda_k^3 \muint \phiint \zeta_k^3}.
\end{aligned}
\label{eq:barotrope}
\end{equation}
Using this expression, we calculate the total potential $U_{\rm pu}$ on the surface
of each layer and at the center using Equations \eqref{eq:surface_subsatellite} and
\eqref{eq:interior_subsatellite}~--~\eqref{eq:central_potential}. Since the density
is constant between interfaces, the hydrostatic equilibrium relation,
Eqn. \eqref{eq:hydrostatic} is trivially integrated to obtain the pressure at the
bottom of the $j$th layer.
\begin{equation}
\begin{aligned}
    P_{j,{\rm pu}} = P_{j-1,{\rm pu}} + \rho_{j-1,{\rm pu}}
    (U_{j,{\rm pu}}-U_{j-1,{\rm pu}})
\end{aligned}
\end{equation}

After obtaining a converged hydrostatic-equilibrium model for N spheroids with the
above array using the initial density profile $\delta_j$, one calculates the arrays
$U_{j,{\rm pu}}$ and $P_{j,{\rm pu}}$. Next, one calculates an array of desired
densities 
\begin{equation}
\begin{aligned}
    \rho_{j,{\rm pu},{\rm desired}} = \rho\left(\frac{1}{2}(P_{j+1}+P_j)\right),
\end{aligned}
\label{eq:rhodesired}
\end{equation}
where $\rho(P)$ is the inverse of the adopted barotrope $P(\rho)$.  Finding the
difference between the desired densities of subsequent layers then gives a new array
of $\delta_j$ for use in the next iteration. In our implementation, it is also
necessary to scale these densities by a constant factor to obtain the correct total
mass of the CMS model. 

Self-gravity from the model's rotational and tidal deformation will cause a small
change in the density profile from that expected for a spherical body.  In practice,
only relatively large changes in the shape of the body will cause a significant
deviation in the density profile. Since $q_{\rm rot} \gg q_{\rm tid}$, the influence
of rotation dominates the shape of the body. For this reason, we can use an
axisymmetric, rotation-only model as described in \citet{hubbard2013} to find a
converged density structure for a given barotrope and specified $q_{\rm rot}$, and
then perform a single further iteration with tides added to find the hydrostatic
solution for that density profile.  Because the tide-induced density changes are very
small, it is unnecessary to iterate with Eqn.  \eqref{eq:rhodesired} to relax the
configuration further for the triaxial figure. Converging the density-pressure
profile to a prescribed barotrope and a fully triaxial figure with relatively large
$q_{\rm tid}$ is significantly more computationally expensive, and is irrelevant to
any giant planet in our planetary system.

\section{Comparison with test cases} \label{tests}

\subsection{Single Maclaurin spheroid} \label{maclaurin}

The well-known special case of a single constant-density Maclaurin spheroid is an
important test, because it has a closed form, analytical solution to the theory of
figures \citep{tassoul2015}. In equilibrium, the Maclaurin spheroid will have an
ellipsoidal shape. In the limit of a low-amplitude tidal perturbation and zero
rotation, the love number for all permitted $n$ is
\begin{equation}
\begin{aligned}
    k_n = \frac{3}{2(n-1)}
\end{aligned}
\label{eq:mac_kn}
\end{equation}
\citep{munk2009}. 

\begin{figure}[h!]  
  \centering
    \includegraphics[width=1.0\textwidth]{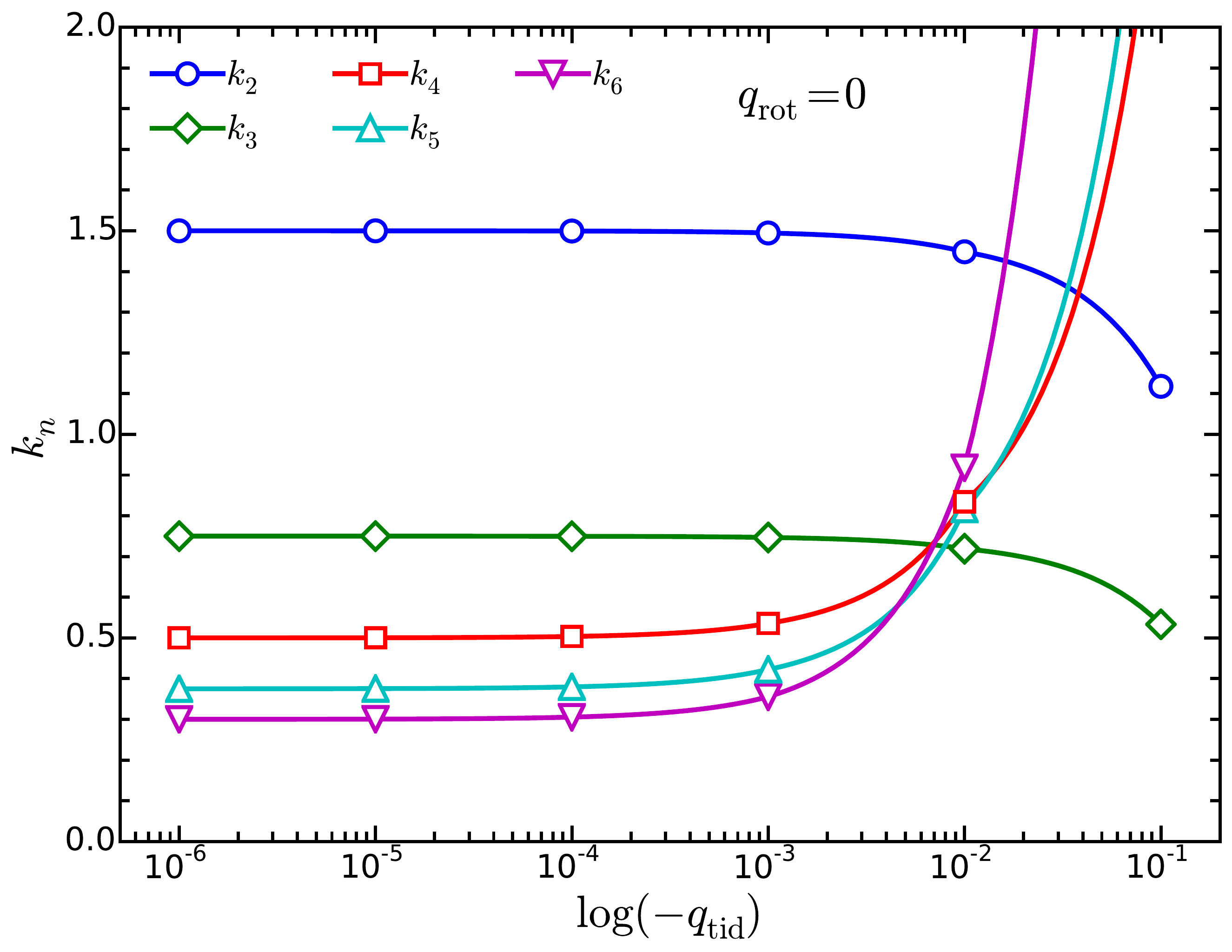}
\caption{ The effect of tidal perturbation strength on the tidal love numbers of a
non-rotating Maclaurin spheroid up to order 6. The love numbers $k_n$ are degenerate
with respect to $m$. The orbital radius is taken to be that of Tethys.}
\label{fig:maclaurin_tidal_only}
\end{figure}

From our simulation results, we calculate the love numbers as
\begin{equation}
\begin{aligned}
    k_{nm} = -\frac{2}{3}\frac{(n+m)!}{(n-m)!}\frac{C_{nm}}{P_n^m(0)q_{\rm tid}}
    \left( \frac{a}{R} \right)^{2-n}.
\end{aligned}
\label{eq:kn}
\end{equation}
For simulations with finite $q_{\rm tid}$ and $q_{\rm rot}=0$, we find our calculated
$k_{nm}$ to be degenerate with $m$ in accordance with the analytical result. For a
given value of $n$, 
\begin{equation}
    k_{nm} = \begin{cases}
        0, & n \mathrm{~and~} m \mathrm{~opposite~parity} \\
        \mathrm{const}, & n \mathrm{~and~} m \mathrm{~same~parity}. \\
    \end{cases}
    \label{eq:degenerate}
\end{equation}
Figure \ref{fig:maclaurin_tidal_only} shows the calculated $k_n$ for the non-rotating
Maclaurin spheroid as a function of $q_{\rm tid}$ up to order $n=6$, with $R/a$ taken
to be that for Tethys and Saturn. For a small tidal perturbation, we find that $k_n$
approaches the analytical result of Eqn. \eqref{eq:mac_kn}. Conversely, as $q_{\rm
rot}$ approaches unity from below, the love numbers diverge, with $k_n$ decreasing
for $n\leq3$ and increasing for $n>3$. The departure from the analytical solution
becomes significant ($\left|\Delta k_n\right| > 0.1$) for $-q_{\rm tid}>10^{-3}$,
whereas for values representative of the largest Saturnian satellites, $k_2$ matches
the analytic value to within our numerical precision.

\begin{figure}[h!]  
  \centering
    \includegraphics[width=1.0\textwidth]{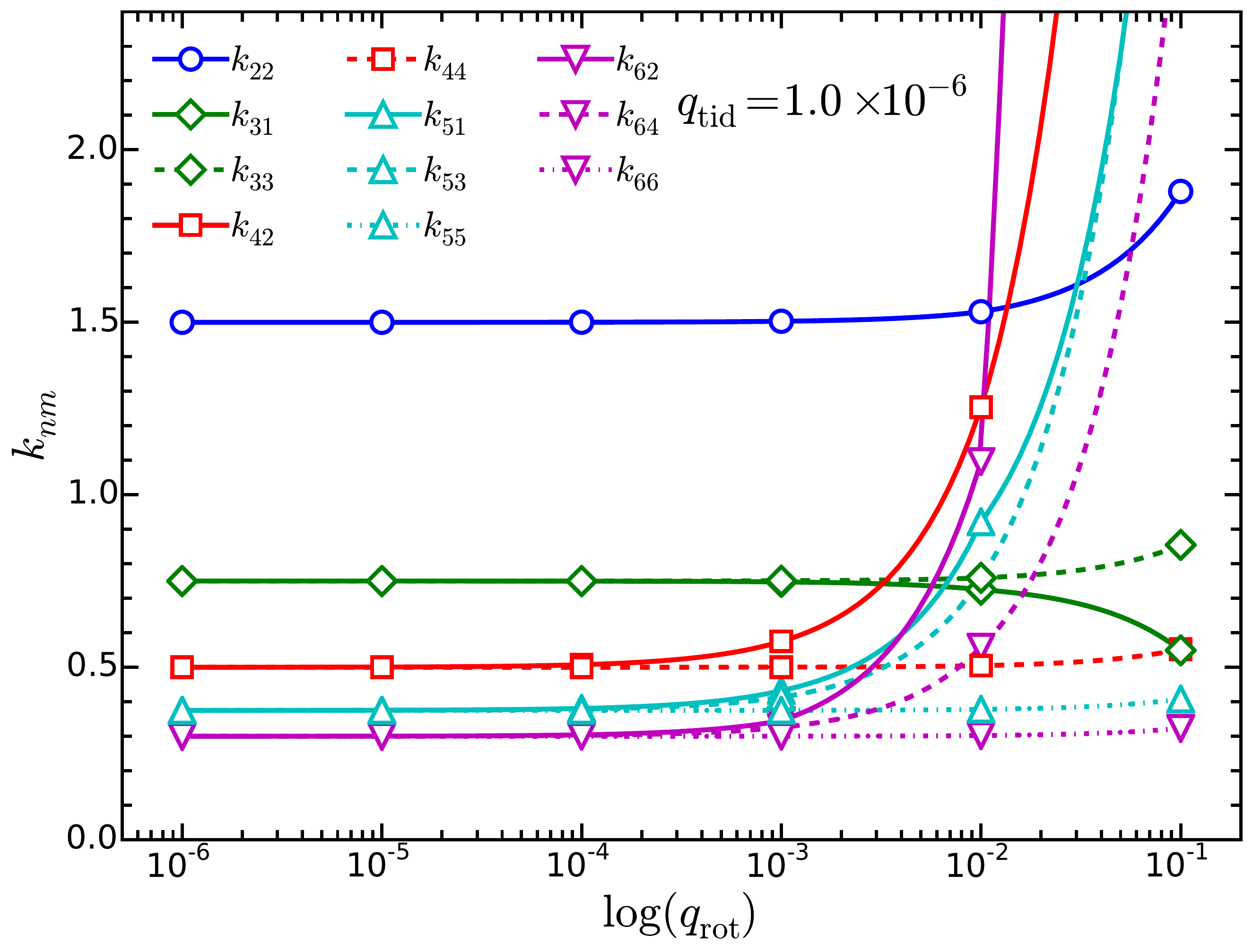}
\caption{ The effect of rotation rate on the tidal love numbers of Maclaurin spheroid
    up to order 6. The $k_{nm}$ for a given $n$ are found to split at high rotation
    rates. $q_{\rm tid}$ is kept constant at $1.0\times10^{-6}$, and the orbital
radius is taken to be that of Tethys.}
\label{fig:maclaurin_rotation_effect}
\end{figure}

In general, the tidal response of a gas giant planet will not be a perturbation to a
perfect sphere, but to a spheroidal shape dominated by rotational flattening.
Therefore, simulation of the tidal response in the absence of rotation is not
generally applicable to real gas giants. When we simulate a Maclaurin spheroid with
both finite $q_{\rm rot}$ and $q_{\rm tid}$, we find a different behavior for
$k_{nm}$ as defined by Eqn. \eqref{eq:kn}. Figure \ref{fig:maclaurin_rotation_effect}
shows the calculated $k_{nm}$ for a Maclaurin spheroid with a constant $q_{\rm tid}$
and a variable $q_{\rm rot}$. When the magnitude of $q_{\rm rot}$ is comparable to
$q_{\rm tid}$, the tidal response matches the expected analytical result. However,
for $q_{\rm rot}>10^{-3}$, we can see that the degeneracy of $k_{nm}$ with $m$ is
broken, and all permitted $k_{nm}$ deviate from the expected values. In other words,
Eqn.  \eqref{eq:degenerate} becomes
\begin{equation}
    \begin{cases}
        k_{nm}=0, & n \mathrm{~and~} m \mathrm{~opposite~parity} \\
        k_{nm}\neq\mathrm{const}, & n \mathrm{~and~} m \mathrm{~same~parity}, \\
    \end{cases}
    \label{eq:nondegenerate}
\end{equation}
and all permitted $k_{nm}$ deviate from the expected values. We also note that these
deviations become pronounced earlier for the higher order $n$.

\subsection{Two concentric Maclaurin Spheroids} \label{two_layer}

Proceeding to more complicated interior structures has proved challenging for
analytical or semi-analytical methods. Even the next simplest model with two
constant-density layers does not have a closed form solution for arbitrary order $n$.
\citet{folonier2015} present an extension of Clairaut theory for a multi-layer planet
under the approximation that the level surfaces are perfect ellipsoids. Under this
approximation, they derive an analytic solution for the distortion in response to a
tidal perturbation only. This yields an expression for $k_2$ as a function of two
ratios of properties of the two layers, $a_1/a$ and $\rho_0/\rho_1$. Table
\ref{tab:folonier_table} shows a comparison of our calculated $k_2$ with the analytic
result from \citet{folonier2015} for a selection of parameters spanning a range of
$a_1/a$ and $\rho_0/\rho_1$. All of our results using the CMS method differ from
those using Clairaut theory by less than $10^{-5}$. This provides an important test
of the correctness of the interior potentials used in our approach. It also indicates
that ellipsoids, while not exact, are a very good approximation for the degree 2
tidal response shape in the limit of very small $q_{\rm tid}$, and $q_{\rm rot}=0$. 

\subsection{Polytrope of index unity} \label{polytrope}

The polytrope of index unity defines a more realistic barotrope that also lends
itself to semi-analytic analyses. It corresponds to the relation
\begin{equation}
    P = K\rho^2
    \label{eq:poly}
\end{equation}
where the polytropic constant $K$ can be chosen to match the planet's physical
parameters. For a nonrotating $n=1$ polytrope, the density distribution is given by
\begin{equation}
    \rho = \rho_c \frac{\sin \pi \lambda}{\pi \lambda}
    \label{eq:poly_rho}
\end{equation}
where $\rho_c$ is the density at the center of the planet.  To obtain the first
approximation of $\delta_j$, we differentiate Eqn. \eqref{eq:poly_rho} by $\lambda$:
\begin{equation}
    \frac{d(\rho/\rho_c)}{d\lambda} = \frac{\cos \pi \lambda}{\lambda} 
    - \frac{\sin \pi \lambda}{\pi \lambda^2}.
\end{equation}
We then correct this profile to be consistent with the given $q_{\rm rot}$ via the
method introduced in Section \ref{barotrope}. Scaling the densities to maintain the
total mass of the planet has a straightforward interpretation for a polytropic
barotrope, as it is equivalent to changing $K$.

For the Maclaurin spheroid the lowest degree love number was 
\begin{equation}
    k_2 = \frac{3}{2}.
\end{equation}
Considering only the linear response to a purely rotational perturbation, we define a
general degree 2 linear response parameter $\Lambda_2$ as 
\begin{equation}
    J_2 = \Lambda_2 q_{\rm rot}.
    \label{eq:linear_response}
\end{equation}
Whereas $\Lambda_2=1/2$ for the Maclaurin spheroid, for the polytrope of index unity
the analytic result is \citep{hubbard1975} 
\begin{equation}
\begin{aligned}
    \Lambda_2 = \left( \frac{5}{\pi^2} - \frac{1}{3} \right).
\end{aligned}
\end{equation}
Considering linear response only, one finds in general
\begin{equation}
\begin{aligned}
    k_2 = 3 \Lambda_2 ,
\end{aligned}
 \label{eq:k2_lambda2}
\end{equation}
valid in the limit $q_{\rm rot} \ll 1$ and $q_{\rm tid} \ll 1$, for any barotrope in
hydrostatic equilibrium.  Thus, for the polytrope of index unity in this limit,
\begin{equation}
\begin{aligned}
    k_2 = \frac{15}{\pi^2} - 1 = 0.519817755.
    \label{eq:poly_estimate}
\end{aligned}
\end{equation}
We compare this to a CMS simulation of the $n=1$ polytrope model with 128 layers,
$q_{\rm rot}=0$, $q_{\rm tid}=10^{-6}$, and Tethys' $R/a$. The simulation results
agree with the expected relation $J_2=2C_{22}$ to numerical precision, and yield
$k_2=0.519775$. This provides a test of the multi-layer CMS approach subject to a
tidal-only perturbation.  The CMS result matches our Eqn. \eqref{eq:poly_estimate}
benchmark to better than the precision with which we could measure this parameter
using the \textit{Juno} spacecraft.  The small difference can be attributed to
approximation of a continuous polytrope by 128 layers in the CMS simulation.
\citet{wisdom2016} (Eqn.  15) show the relative discretization error of a CMS
polytrope model to be $\sim 10^{-3}$ for $N=128$, roughly consistent with our
calculated difference. 

\begin{figure}[h!]  
  \centering
    \includegraphics[width=1.0\textwidth]{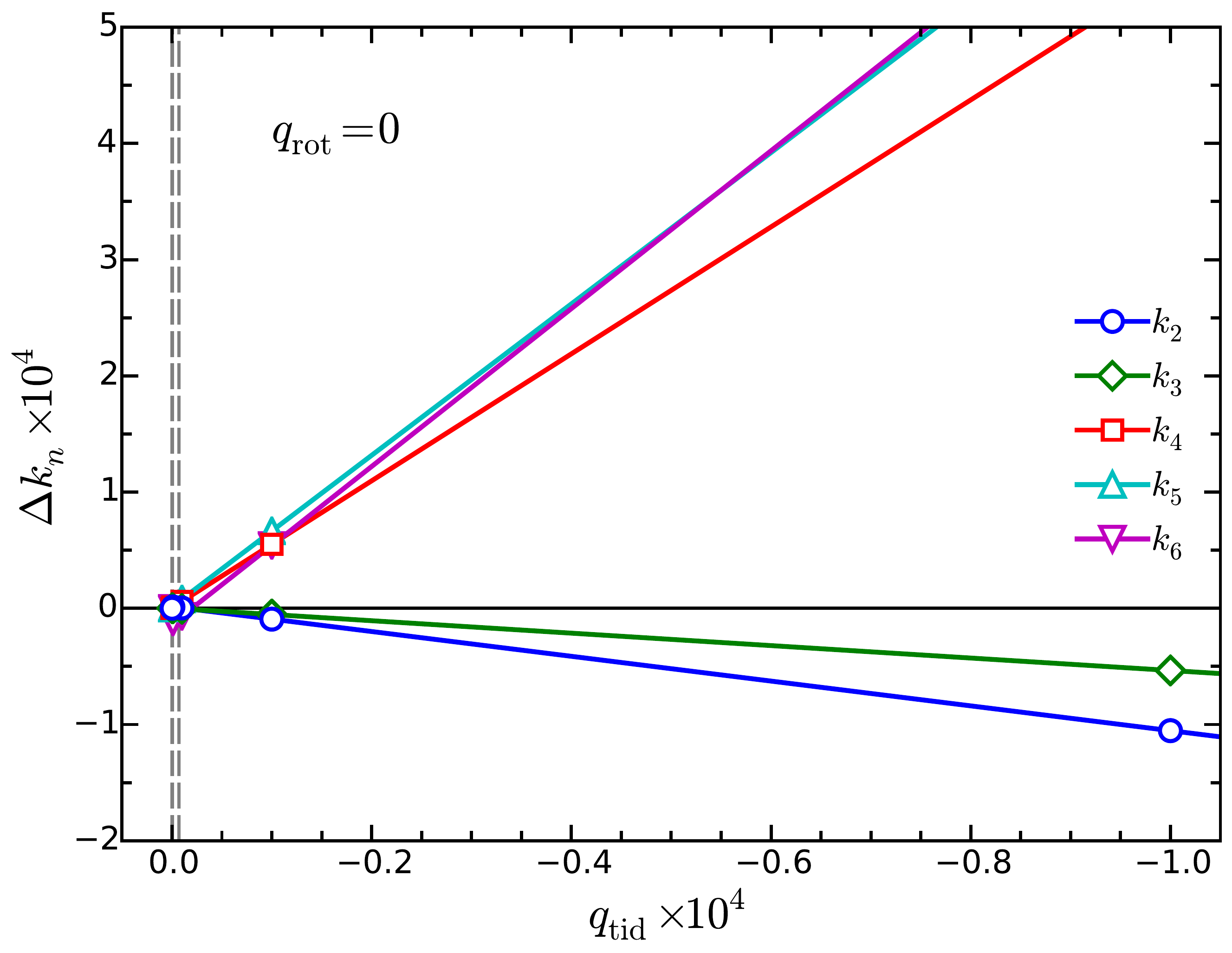}
\caption{ The effect of tidal perturbation strength on the tidal love numbers of a
    non-rotating planet with an $N=1$ polytrope equation of state, up to order 6.
    $\Delta k_n$ is the shift in love number $k_n$ from the limit of low $q{\rm
    tid}$. The love numbers $k_n$ are degenerate with respect to $m$. The orbital
    radius is taken to be that of Tethys. The vertical, dashed gray lines show $q_{\rm tid}$
    for Tethys-Saturn and Io-Jupiter.}
\label{fig:polytrope_tidal_only}
\end{figure}

Similar to the calculations on the Maclaurin spheroid in Section \ref{maclaurin}, we
performed additional $N=128$ polytrope simulations with finite $q_{\rm tid}$ and
$q_{\rm rot}=0$. Once again, we find our calculated $k_{nm}$ to be degenerate with
$m$ for the tidal-only simulations, in agreement with Eqn. \eqref{eq:degenerate}.
Figure \ref{fig:polytrope_tidal_only} shows the behavior of $k_n$ for $n\leq6$ for
these tidal-only polytrope simulations.  We only present these results up to $q_{\rm
tid}\sim10^{-4}$, because above that value effects of the triaxial shape on the
pressure-density profile would require iterated relaxation to the polytropic
relation, as discussed in Section \ref{barotrope}. We observe that realistic values
for $q_{\rm tid}$ have negligible effect on the tidal response. Even for the
Io-Jupiter system, the effect of finite $q_{\rm tid}$ on $k_{nm}$ is near the
numerical noise level. The general behavior is quite similar to the case of the
single Maclaurin spheroid.  For small tidal perturbations, the polytrope $k_n$
approach values smaller than the Maclaurin spheroid case, with $k_2$ asymptoting to
the analytic limit in Eqn.  \eqref{eq:poly_estimate}.  Similar to the Maclaurin
spheroid, the behavior as $q_{\rm rot}$ increases from zero sees $k_n$ decrease for
$n\leq3$ and increase for $n>3$.  The deviation from the low $q_{\rm tid}$ value is
also less pronounced for the more realistic polytrope density distribution than for
the Maclaurin spheroid. This is to be expected since there is less mass concentrated
in the outer portion of the polytrope model. 

\begin{figure}[h!]  
  \centering
    \includegraphics[width=0.9\textwidth]{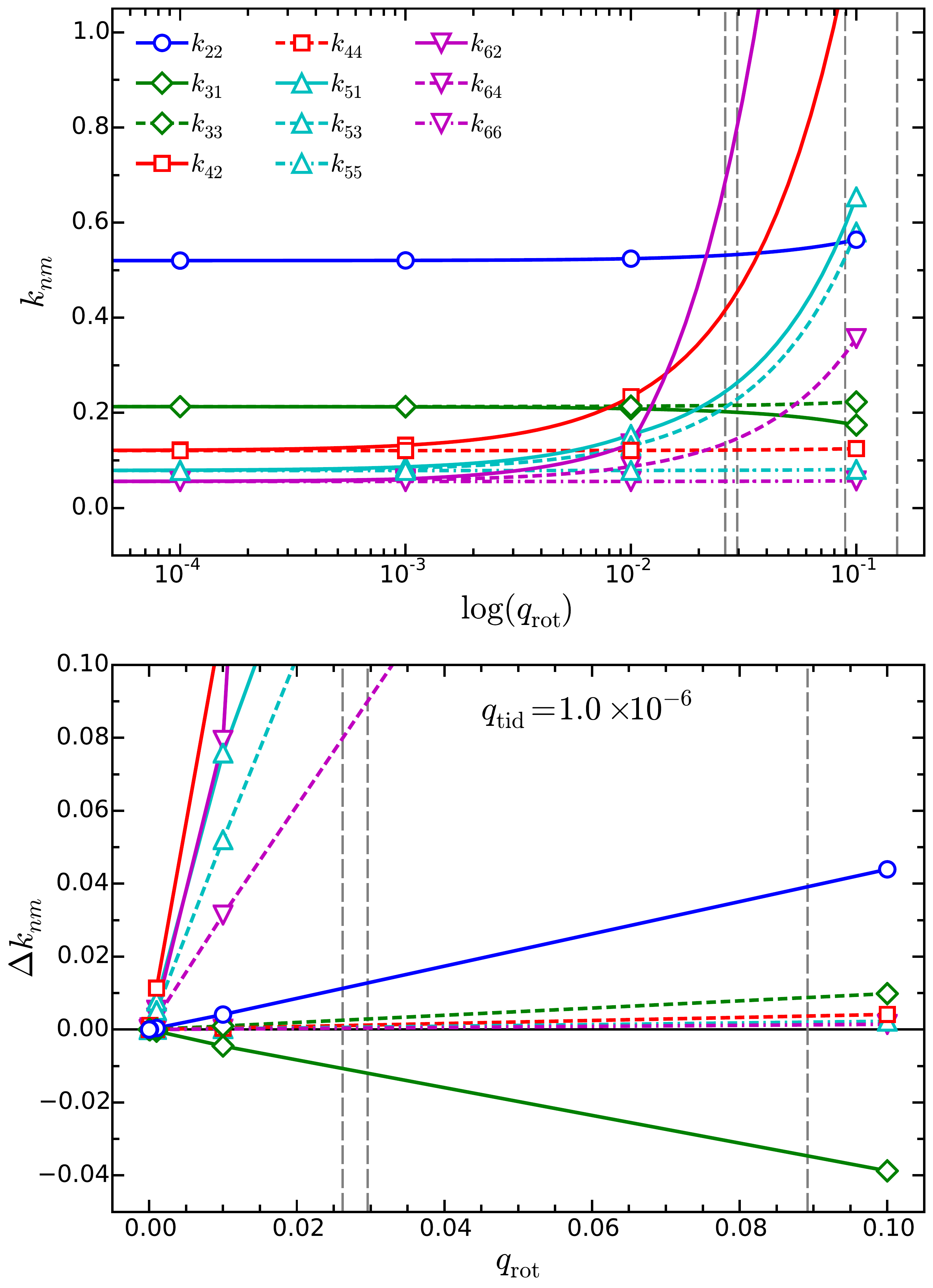}
\caption{ Top: The effect of rotation rate on the tidal love numbers of a planet with an
    $N=1$ polytrope equation of state, up to order 6. The $k_{nm}$ for a given $n$
    are found to split at high rotation rates. $q_{\rm tid}$ is kept constant at
    $1.0\times10^{-6}$, and the orbital radius is taken to be that of Tethys. The
    vertical, dashed gray lines show $q_{\rm rot}$ for Neptune, Uranus, Jupiter and Saturn.
    Bottom: Shift in $k_{nm}$ as a function of $q_{\rm rot}$ on a linear scale.}
\label{fig:polytrope_rotation_effect}
\end{figure}

Figure \ref{fig:polytrope_rotation_effect} shows the effect of variable $q_{\rm rot}$
on polytrope models with constant $q_{\rm tid}$. Once again, we find that $k_{nm}$
degeneracy with respect to $m$ breaks, in agreement with Eqn.
\eqref{eq:nondegenerate}, as $q_{\rm rot}$ increases. Although the splitting of
$k_{nm}$ is somewhat diminished from the single Maclaurin spheroid results, the
deviations are still significant at large values of $q_{\rm rot}\sim 10^{-2}$
consistent with the rapidly-rotating gas giants. The shift in $k_{nm}$ shows a nearly
linear increase in magnitude with increasing $q_{\rm rot}$, with potentially
observable increases in $k_2$ for both the ice giant and gas giant planets. The
general behavior of $k_{nm}$ is very similar between these tests with two very
different density profiles. The relative magnitudes and directions of all $k_{nm}$ up
to $n=6$ are similar between the two cases. This indicates that the effect should be
ubiquitous in all fast-spinning liquid bodies, and relatively insensitive to the
density profile of the planet.

\section{Saturn's tidal response} \label{saturn}

\subsection{Saturn interior models}

\citet{lainey2016} present the first determination of the love number $k_2$ for a gas
giant planet using a dataset of astrometric observations of Saturn's coorbital moons.
Their observed value $k_2=0.390 \pm 0.024$ is much larger than the theoretical
prediction of 0.341 by \citet{gavrilov1977}. Here we present calculations suggesting
that the enhancement of Saturn's $k_2$ is the result of the influence of the planet's
rapid rotation, rather than evidence for a nonstatic tidal response or some other
breakdown of the hydrostatic theory.

For the purposes of this calculation, we use two relatively simple models for
Saturn's interior structure, fitted to physical parameters determined by the
\textit{Voyager} and \textit{Cassini} spacecraft. Table \ref{tab:saturn_params}
summarizes the physical parameters used in our models. We fit our models to minimize
the difference in zonal harmonics from those determined from \textit{Cassini}
\citep{Jacobson2006}.  We consider two different internal rotation rates based on
magnetic field measurements from \textit{Voyager} \citep{desch1981} and
\textit{Cassini} \citep{giampieri2006}, which lead to two different values of $q_{\rm
rot}$. 

In principle, the tidal response of a heterogeneous body will also be different for
satellites with different sizes and orbital parameters. To address this, we also
consider the effect of two major satellites, Tethys and Dione, with different values
for $q_{\rm tid}$ and $R/a$ \citep{archinal2011}. These two satellites, along with
their respective coorbital satellites, were used in the determination of $k_2$ by
\citet{lainey2016}.

\begin{figure}[h!]  
  \centering
    \includegraphics[width=1.0\textwidth]{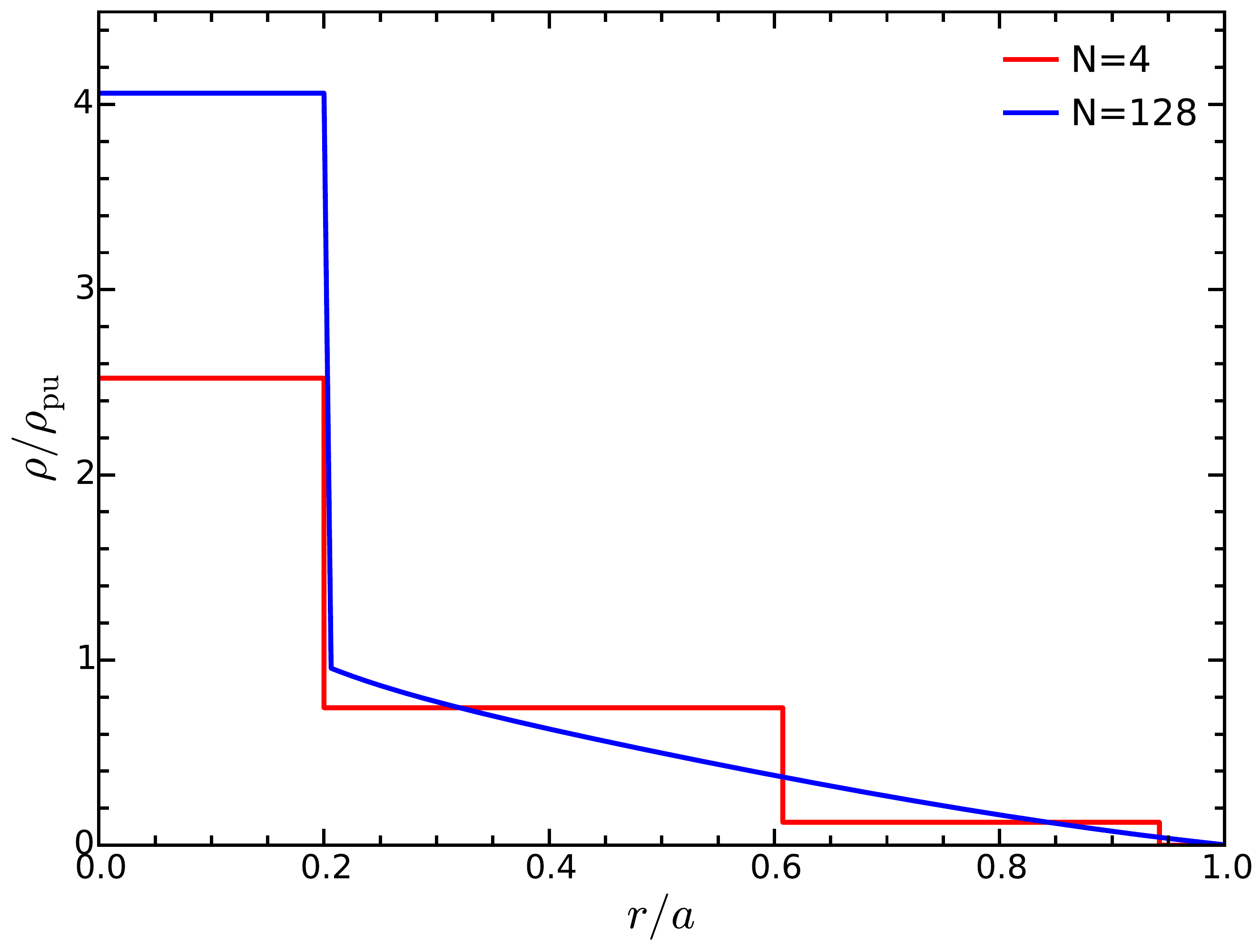} \caption{ Density
    structure of simple Saturn models. The blue curve shows an $N=128$ model having a
    dense core within $r=0.2a$ and a polytropic outer envelope. The red curve shows
    an $N=4$ model with the same core constraints. Both models have densities
    adjusted to match $J_2$ measured by \textit{Cassini} \citep{Jacobson2006}.}
\label{fig:density_structure}
\end{figure}

For the interior density profile, our first model assumes a constant-density core
surrounded by a polytropic envelope following Eqn. \eqref{eq:poly}. We constrain the
radius of the core to be $a_{\rm core}/a=0.2$, leaving the mass $m_{\rm core}/M$ as a
parameter which is adjusted to match the observed Saturn $J_2$.  The fitted model
using the Voyager rotation period matches both $J_2$ and $J_4$ to within the error
bars, but with the Cassini rotation period it matches only $J_2$.  In hydrostatic
equilibrium, the two different rotation rates lead to differences in shape of
equipotential surfaces and, therefore, also to different best fits to $m_{\rm
core}/M$. The envelope polytrope is scaled in order to maintain $M$. Figure
\ref{fig:density_structure} shows the density profile of one such model. We consider
a model with a total of 128 layers, for which the CMS model has a discretization
error \citep{wisdom2016} smaller than uncertainty in the observations of Saturn's
$k_2$.

Our second model has only four spheroids ($N=4$), also depicted in Figure
\ref{fig:density_structure}, with densities and radii adjusted to yield agreement
with both observed $J_2$ and observed $J_4$ as given in Table
\ref{tab:saturn_params}.

These two simple models, while not particularly realistic, capture the major features
of Saturn's internal structure. It is well established that the details of Saturn's
internal structure are largely degenerate, with a wide range of possible core sizes
and densities adequately matching the few observational constraints
\citep{kramm2011,helled2013,Nettelmann2013}.  The qualitative similarities between
our Maclaurin spheroid and polytrope simulations (Sections \ref{maclaurin} and
\ref{polytrope}) indicate that the rotational enhancement of $k_2$ should be a robust
prediction regardless of the particular details of the interior profile. A comparison
between our polytrope plus core and four layer models provides another test of the
sensitivity of $k_2$ to interior structure.  We do not consider here the influence of
differential rotation \citep{hubbard1982,Kong2013,cao2015,wisdom2016}, which might
have an influence on the gravitational response in comparison to the solid-body
rotation considered here.  However, since the effect of realistic deep flow patterns
on the low order zonal harmonics is small \citep{cao2015}, we expect that they would
cause negligible further changes in the rotational enhancement of $k_2$.

\subsection{Calculated $k_2$ for Saturn}

\begin{figure}[h!]  
  \centering
    \includegraphics[width=1.0\textwidth]{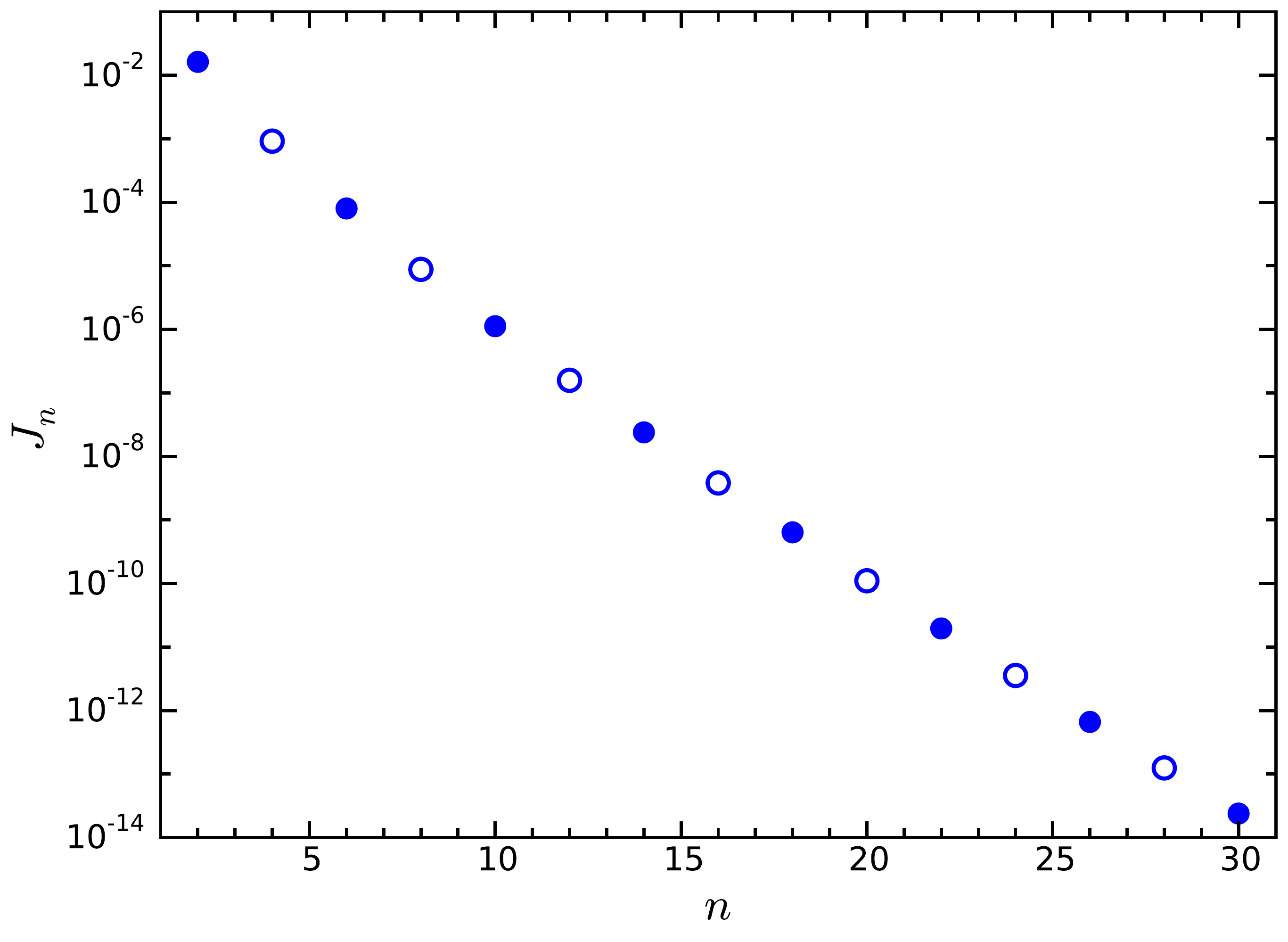}
\caption{ The zonal harmonics $J_n$ for the \textit{Cassini} Saturn model. Positive
values are shown as filled and negative  as empty.}
\label{fig:saturn_zonal}
\end{figure}

We take our baseline model to be the $N=128$ CMS core plus polytrope model with
physical parameters fitted to \textit{Cassini} observations. Figure
\ref{fig:saturn_zonal} shows the calculated zonal harmonics $J_n$ up to order $n=30$.
The even $J_n$ decrease smoothly in magnitude with increasing $n$, with the slope
decreasing at higher $n$.  $J_n$ is negative when $n$ is divisible by 4, and positive
otherwise.  The calculated $J_n$ are essentially indistinguishable from those
calculated for the rotation only case with the same $q_{\rm rot}$, as is expected
given $q_{\rm rot} \gg q_{\rm tid}$.

\begin{figure}[h!]  
  \centering
    \includegraphics[width=1.0\textwidth]{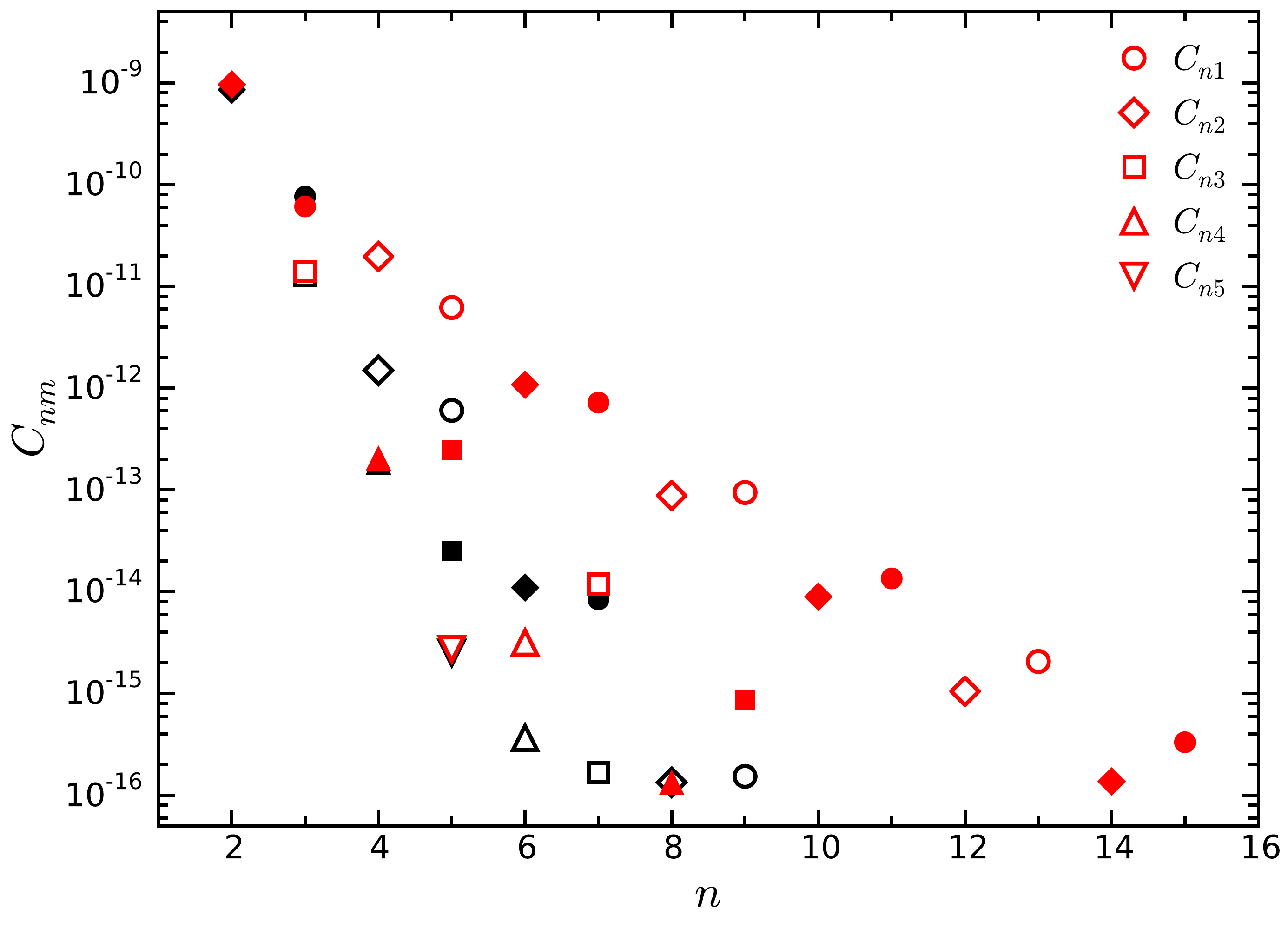}
\caption{ In red, the tesseral harmonics $C_{nm}$ for the \textit{Cassini} Saturn
model. In black, $C_{nm}$ for the same density profile and same value of $q_{\rm
tid}$, but with $q_{\rm rot}=0$. Positive values are shown as filled and negative as
empty.} \label{fig:saturn_tesseral}
\end{figure}

Figure \ref{fig:saturn_tesseral} shows the magnitude of $C_{nm}$ for the core plus
polytrope model with \textit{Cassini} rotation. Changing the number of layers,
satellite parameters or the rotation rate to the \textit{Voyager} value leads to a
shift in the values, but the relative magnitudes and signs of $C_{nm}$ remain
approximately the same. In the same figure, we also compare the $C_{nm}$ for a
non-rotating planet having the same density profile $\rho( \lambda_i)$. Here we see
significant shifts in the magnitudes $C_{nm}$, although the signs remain the same.
For the rotating model, $C_{nm}$ is similar for most points where $n=m$, but with
magnitudes significantly larger when $m<n$. The only exception to this trend is
$C_{31}$ which is lower for the rotating model. These results are all broadly
consistent with the splitting of $k_{nm}$ observed for the polytrope in Section
\ref{polytrope}.

Table \ref{tab:saturn_results} summarizes our calculated values for $k_2$ for 5
different models. The identifying labels ``Cassini'' and ``Voyager'' use the observed
rotation rate from \citet{Jacobson2006}, and \citet{desch1981} respectively, while
``non-rotating'' is a model with $q_{\rm rot}=0$. The ``non-rotating'' model uses the
same ``Cassini'' density profile, meaning that its density-pressure profile has not
been relaxed to be in equilibrium for zero rotation.  It does, however, allow us to
quantify the effect of rotation on the tidal response by comparison with the
``Cassini'' model. ``Tethys'' and ``Dione'' refer to models with the satellite
parameters $q_{\rm tid}$ and $R/a$ corresponding to those satellites, whereas
``no~tide'' is an analogous model with finite $q_{\rm rot}$ only. ``$N=128$'' uses
the polytrope outer envelope with constant density inner core, whereas ``$N=4$'' is
the model which independently adjusts layer densities to match the observed $J_2$ and
$J_4$. 

Each of the rotating models yields a calculated $k_2$ value matching the observation
of \citet{lainey2016} within their error bars. We find that the difference between
the $k_2$ values associated with the satellites Tethys and Dione is $\sim$0.0003,
well below the current sensitivity limit. Using the $\sim$2.5\% higher ``Voyager''
rotation rate leads to a decrease of $\sim$0.01 in $k_2$. 

In Table \ref{tab:saturn_results}, we also show the calculated $J_2$, $J_4$ and $J_6$
following the convergence of the gravitational field in response to the tidal
perturbation. For the core plus polytrope model, the rotation rate from
\textit{Voyager} is more consistent with the $J_4$ and $J_6$ from
\citet{Jacobson2006}. This doesn't necessarily mean that the \textit{Voyager}
rotation rate is more correct, just that it allows a better fit for our simplified
density model. Nonetheless, our fitted gravitational moments are much closer to each
other than to those from the pre-\textit{Cassini} model of \citet{gavrilov1977}.

In comparison to the other models, the outlier is the non-rotating model, which
underestimates the $k_2$ by $\sim9.4$\% compared to a rotating body with the same
density distribution.  This calculated enhancement accounts for most of the
difference between the observation of $k_2=0.390 \pm .024$ \citep{lainey2016} and the
classical theory result of 0.341 \citep{gavrilov1977}. We attribute our non-rotating
model's larger $k_2$ to our different interior model which matches more recent
constraints on Saturn's zonal gravitational moments $J_2$--$J_6$.

In addition to the difference in $k_2$, the non-rotating model also predicts slightly
different tidal components of the zonal gravitational moments. Finding the difference
in values between the ``no tide'' model and the analogous tidal model yields
$J_{2,{\rm tid}}=1.7254\times10^{-10}$, $J_{4,{\rm tid}}=-2.732\times10^{-11}$ and
$J_{6,{\rm tid}}=4.14\times10^{-12}$, which are different than calculated zonal
moments for the ``non-rotating'' model.

It may be initially surprising that the four-layer model yields a $k_2$ value only
$\sim$0.0007 different than the polytrope model. The two models represent two very
different density structures that lead to similar low-order zonal harmonics. The fact
these two models are indistinguishable by their $k_2$ suggests that the tidal
response of Saturn is only a weak function of the detailed density structure within
the interior of the planet. This behavior can be understood by referring to Eqn.
\eqref{eq:k2_lambda2}, which shows that to lowest order, $k_2$ and $\Lambda_2$
contain the same information about interior structure.  This statement is not true
when we include a nonlinear response to rotation and tides.  Thus, future
high-precision measurements of the $k_{nm}$ of jovian planets, say to better than
$0.1\%$, will be useful for constraining basic parameters such as the interior
rotation rate of the planet, and may help to break the current degeneracy of interior
density profiles.  The theory presented in this paper is intended to match the
anticipated precision of such future measurements.

\section{Summary} \label{summary}

The CMS method for calculating a self-consistent shape and gravitational field of a
static liquid planet has been extended to include the effect of a tidal potential
from a satellite. This is expected to represent the largest contribution to the
low-order tesseral harmonics measured by \textit{Juno} and future spacecraft studies
of the gas giants. This approach has been benchmarked against analytical results for
the tidal response of the Maclaurin spheroid, two constant density layers, and the
polytrope of index unity. 

We highlight for the first time the importance of the high rotation rate on the tidal
response of the gas giants. CMS simulations of the tidal response on bodies with
large rotational flattening show significant deviation in the tesseral harmonics of
the gravitational field as compared to simulations without rotation. This includes
splitting of the love numbers into different $k_{nm}$ for any given order $n>2$.
Meanwhile, it leads to an observable enhancement in $k_2$ compared to a non-rotating
model.

This rotational enhancement of the $k_2$ love number for a simplified interior model
of Saturn agrees with the recent observational result \citep{lainey2016}, which found
$k_2$ to be much higher than previous predictions. Our predicted values of $k_2$ are
robust for reasonable assumptions of interior structure, rotation rate and satellite
parameters.  The \textit{Juno} spacecraft is expected to measure Jupiter's
gravitational field to sufficiently high precision to measure lower order tesseral
components arising from Jupiter's large moons, and we predict an analogous rotational
enhancement of $k_2$ for Jupiter.  Our high-precision tidal theory will be an
important component of the search for non-hydrostatic terms in Jupiter's external
gravity field.

\section*{Acknowledgments}

This work was supported by NASA's \textit{Juno} project. Sean Wahl and Burkhard
Militzer acknowledge support the National Science Foundation (astronomy and
astrophysics research grant 1412646). We thank Isamu Matsuyama for helpful
discussions regarding classical tidal theory.

\section*{References}

\bibliography{saturn}{}

\begin{thebibliography}{25}
\expandafter\ifx\csname natexlab\endcsname\relax\def\natexlab#1{#1}\fi
\providecommand{\url}[1]{\texttt{#1}}
\providecommand{\href}[2]{#2}
\providecommand{\path}[1]{#1}
\providecommand{\DOIprefix}{doi:}
\providecommand{\ArXivprefix}{arXiv:}
\providecommand{\URLprefix}{URL: }
\providecommand{\Pubmedprefix}{pmid:}
\providecommand{\doi}[1]{\href{http://dx.doi.org/#1}{\path{#1}}}
\providecommand{\Pubmed}[1]{\href{pmid:#1}{\path{#1}}}
\providecommand{\bibinfo}[2]{#2}
\ifx\xfnm\relax \def\xfnm[#1]{\unskip,\space#1}\fi
\bibitem[{Archinal et~al.(2011)Archinal, A'Hearn, Bowell, Conrad, Consolmagno,
  Courtin, Fukushima, Hestroffer, Hilton, Krasinsky, Neumann, Oberst,
  Seidelmann, Stooke, Tholen, Thomas and Williams}]{archinal2011}
\bibinfo{author}{Archinal, B.A.}, \bibinfo{author}{A'Hearn, M.F.},
  \bibinfo{author}{Bowell, E.}, \bibinfo{author}{Conrad, A.},
  \bibinfo{author}{Consolmagno, G.J.}, \bibinfo{author}{Courtin, R.},
  \bibinfo{author}{Fukushima, T.}, \bibinfo{author}{Hestroffer, D.},
  \bibinfo{author}{Hilton, J.L.}, \bibinfo{author}{Krasinsky, G.A.},
  \bibinfo{author}{Neumann, G.}, \bibinfo{author}{Oberst, J.},
  \bibinfo{author}{Seidelmann, P.K.}, \bibinfo{author}{Stooke, P.},
  \bibinfo{author}{Tholen, D.J.}, \bibinfo{author}{Thomas, P.C.},
  \bibinfo{author}{Williams, I.P.}, \bibinfo{year}{2011}.
\newblock \bibinfo{title}{{Report of the IAU Working Group on Cartographic
  Coordinates and Rotational Elements: 2009}}.
\newblock \bibinfo{journal}{Celest. Mech. Dyn. Astron.} \bibinfo{volume}{109},
  \bibinfo{pages}{101--135}.
\newblock \DOIprefix\doi{10.1007/s10569-010-9320-4}.
\bibitem[{Cao and Stevenson(2015)}]{cao2015}
\bibinfo{author}{Cao, H.}, \bibinfo{author}{Stevenson, D.J.},
  \bibinfo{year}{2015}.
\newblock \bibinfo{title}{{Gravity and Zonal Flows of Giant Planets: From the
  Euler Equation to the Thermal Wind Equation}} ,
  \bibinfo{pages}{1--9}\URLprefix \url{http://arxiv.org/abs/1508.02764},
  \href{http://arxiv.org/abs/1508.02764}{\tt arXiv:1508.02764}.
\bibitem[{Desch and Kaiser(1981)}]{desch1981}
\bibinfo{author}{Desch, M.D.}, \bibinfo{author}{Kaiser, M.L.},
  \bibinfo{year}{1981}.
\newblock \bibinfo{title}{{Voyager measurement of the rotation period of
  Saturn's magnetic field}}.
\newblock \bibinfo{journal}{Geophys. Res. Lett.} \bibinfo{volume}{8},
  \bibinfo{pages}{253--256}.
\newblock \DOIprefix\doi{10.1029/GL008i003p00253}.
\bibitem[{Folonier et~al.(2015)Folonier, Ferraz-Mello and
  Kholshevnikov}]{folonier2015}
\bibinfo{author}{Folonier, H.}, \bibinfo{author}{Ferraz-Mello, S.},
  \bibinfo{author}{Kholshevnikov, K.V.}, \bibinfo{year}{2015}.
\newblock \bibinfo{title}{{The flattenings of the layers of rotating planets
  and satellites deformed by a tidal potential}}.
\newblock \bibinfo{journal}{Celest. Mech. Dyn. Astron.} \bibinfo{volume}{122},
  \bibinfo{pages}{183--198}.
\newblock \URLprefix \url{http://arxiv.org/abs/1503.08051},
  \DOIprefix\doi{10.1007/s10569-015-9615-6},
  \href{http://arxiv.org/abs/1503.08051}{\tt arXiv:1503.08051}.
\bibitem[{Gavrilov and Zharkov(1977)}]{gavrilov1977}
\bibinfo{author}{Gavrilov, S.V.}, \bibinfo{author}{Zharkov, V.N.},
  \bibinfo{year}{1977}.
\newblock \bibinfo{title}{{Love numbers of the giant planets}}.
\newblock \bibinfo{journal}{Icarus} \bibinfo{volume}{32},
  \bibinfo{pages}{443--449}.
\newblock \URLprefix
  \url{http://linkinghub.elsevier.com/retrieve/pii/001910357790015X},
  \DOIprefix\doi{10.1016/0019-1035(77)90015-X}.
\bibitem[{Giampieri et~al.(2006)Giampieri, Dougherty, Smith and
  Russell}]{giampieri2006}
\bibinfo{author}{Giampieri, G.}, \bibinfo{author}{Dougherty, M.K.},
  \bibinfo{author}{Smith, E.J.}, \bibinfo{author}{Russell, C.T.},
  \bibinfo{year}{2006}.
\newblock \bibinfo{title}{{A regular period for Saturn's magnetic field that
  may track its internal rotation.}}
\newblock \bibinfo{journal}{Nature} \bibinfo{volume}{441},
  \bibinfo{pages}{62--64}.
\newblock \DOIprefix\doi{10.1038/nature04750}.
\bibitem[{Helled and Guillot(2013)}]{helled2013}
\bibinfo{author}{Helled, R.}, \bibinfo{author}{Guillot, T.},
  \bibinfo{year}{2013}.
\newblock \bibinfo{title}{{Interior Models of Saturn: Including the
  Uncertainties in Shape and Rotation}}.
\newblock \bibinfo{journal}{Astrophys. J.} \bibinfo{volume}{767},
  \bibinfo{pages}{113}.
\newblock \URLprefix
  \url{http://stacks.iop.org/0004-637X/767/i=2/a=113?key=crossref.045d858be83734acdc0600277a318377},
  \DOIprefix\doi{10.1088/0004-637X/767/2/113}.
\bibitem[{Hubbard(1975)}]{hubbard1975}
\bibinfo{author}{Hubbard, W.}, \bibinfo{year}{1975}.
\newblock \bibinfo{title}{{Gravitational field of a rotating planet with a
  polytropic index of unity}}.
\newblock \bibinfo{journal}{Sov. Astron.} \bibinfo{volume}{18},
  \bibinfo{pages}{621--624}.
\bibitem[{Hubbard(1982)}]{hubbard1982}
\bibinfo{author}{Hubbard, W.}, \bibinfo{year}{1982}.
\newblock \bibinfo{title}{{Effects of differential rotation on the
  gravitational figures of Jupiter and Saturn}}.
\newblock \bibinfo{journal}{Icarus} \bibinfo{volume}{52},
  \bibinfo{pages}{509--515}.
\newblock \URLprefix
  \url{http://linkinghub.elsevier.com/retrieve/pii/0019103582900112},
  \DOIprefix\doi{10.1016/0019-1035(82)90011-2}.
\bibitem[{Hubbard et~al.(2014)Hubbard, Schubert, Kong and Zhang}]{hubbard2014}
\bibinfo{author}{Hubbard, W.}, \bibinfo{author}{Schubert, G.},
  \bibinfo{author}{Kong, D.}, \bibinfo{author}{Zhang, K.},
  \bibinfo{year}{2014}.
\newblock \bibinfo{title}{{On the convergence of the theory of figures}}.
\newblock \bibinfo{journal}{Icarus} \bibinfo{volume}{242},
  \bibinfo{pages}{138--141}.
\newblock \URLprefix
  \url{http://linkinghub.elsevier.com/retrieve/pii/S001910351400428X},
  \DOIprefix\doi{10.1016/j.icarus.2014.08.014}.
\bibitem[{Hubbard(2012)}]{hubbard2012}
\bibinfo{author}{Hubbard, W.B.}, \bibinfo{year}{2012}.
\newblock \bibinfo{title}{{High-Precision Maclaurin-Based Models of Rotating
  Liquid Planets}}.
\newblock \bibinfo{journal}{Astrophys. J.} \bibinfo{volume}{756},
  \bibinfo{pages}{L15}.
\newblock \URLprefix
  \url{http://stacks.iop.org/2041-8205/756/i=1/a=L15?key=crossref.34b95153bc3fdfb844cab51abbdf75d3},
  \DOIprefix\doi{10.1088/2041-8205/756/1/L15}.
\bibitem[{Hubbard(2013)}]{hubbard2013}
\bibinfo{author}{Hubbard, W.B.}, \bibinfo{year}{2013}.
\newblock \bibinfo{title}{{Concentric Maclaurin Spheroid Models of Rotating
  Liquid Planets}}.
\newblock \bibinfo{journal}{Astrophys. J.} \bibinfo{volume}{768},
  \bibinfo{pages}{43}.
\newblock \URLprefix
  \url{http://stacks.iop.org/0004-637X/768/i=1/a=43?key=crossref.a31bd47c857111e805198695ba70780e},
  \DOIprefix\doi{10.1088/0004-637X/768/1/43}.
\bibitem[{Jacobson et~al.(2006)Jacobson, Antresian, Bordi, Criddle, Ionasescu,
  Jones, Mackenzie, Meek, Parcher, Pelletier, Owen, Roth, Roundhill and
  Stauch}]{Jacobson2006}
\bibinfo{author}{Jacobson, R.A.}, \bibinfo{author}{Antresian, P.G.},
  \bibinfo{author}{Bordi, J.J.}, \bibinfo{author}{Criddle, K.E.},
  \bibinfo{author}{Ionasescu, R.}, \bibinfo{author}{Jones, J.B.},
  \bibinfo{author}{Mackenzie, R.a.}, \bibinfo{author}{Meek, M.C.},
  \bibinfo{author}{Parcher, D.}, \bibinfo{author}{Pelletier, F.J.},
  \bibinfo{author}{Owen, W.M.}, \bibinfo{author}{Roth, D.C.},
  \bibinfo{author}{Roundhill, I.M.}, \bibinfo{author}{Stauch, J.R.},
  \bibinfo{year}{2006}.
\newblock \bibinfo{title}{{The gravity field of the Saturnian system from
  staellites observations and spacecraft tracking data}}.
\newblock \bibinfo{journal}{Astrophys. J.} \bibinfo{volume}{132},
  \bibinfo{pages}{2520--2526}.
\newblock \DOIprefix\doi{10.1086/508812}.
\bibitem[{Jeans(2009)}]{jeans1919}
\bibinfo{author}{Jeans, J.H.}, \bibinfo{year}{2009}.
\newblock \bibinfo{title}{{Problems of Cosmology and Stellar Dynamics}}.
\newblock \bibinfo{publisher}{Cambridge University Press}.
\newblock \URLprefix \url{http://dx.doi.org/10.1017/CBO9780511694417}.
\bibitem[{Kaspi(2013)}]{kaspi2013}
\bibinfo{author}{Kaspi, Y.}, \bibinfo{year}{2013}.
\newblock \bibinfo{title}{{Inferring the depth of the zonal jets on Jupiter and
  Saturn from odd gravity harmonics}}.
\newblock \bibinfo{journal}{Geophys. Res. Lett.} \bibinfo{volume}{40},
  \bibinfo{pages}{676--680}.
\newblock \DOIprefix\doi{10.1029/2012GL053873}.
\bibitem[{Kaspi et~al.(2010)Kaspi, Hubbard, Showman and Flierl}]{kaspi2010}
\bibinfo{author}{Kaspi, Y.}, \bibinfo{author}{Hubbard, W.B.},
  \bibinfo{author}{Showman, A.P.}, \bibinfo{author}{Flierl, G.R.},
  \bibinfo{year}{2010}.
\newblock \bibinfo{title}{{Gravitational signature of Jupiter's internal
  dynamics}}.
\newblock \bibinfo{journal}{Geophys. Res. Lett.} \bibinfo{volume}{37},
  \bibinfo{pages}{L01204}.
\newblock \URLprefix \url{http://doi.wiley.com/10.1029/2012GL053873},
  \DOIprefix\doi{10.1029/2009GL041385}.
\bibitem[{Kong et~al.(2013)Kong, Liao, Zhang and Schubert}]{Kong2013}
\bibinfo{author}{Kong, D.}, \bibinfo{author}{Liao, X.}, \bibinfo{author}{Zhang,
  K.}, \bibinfo{author}{Schubert, G.}, \bibinfo{year}{2013}.
\newblock \bibinfo{title}{{Gravitational signature of rotationally distorted
  Jupiter caused by deep zonal winds}}.
\newblock \bibinfo{journal}{Icarus} \bibinfo{volume}{226},
  \bibinfo{pages}{1425--1430}.
\newblock \URLprefix
  \url{http://linkinghub.elsevier.com/retrieve/pii/S0019103513003540},
  \DOIprefix\doi{10.1016/j.icarus.2013.08.016}.
\bibitem[{Kramm et~al.(2011)Kramm, Nettelmann, Redmer and
  Stevenson}]{kramm2011}
\bibinfo{author}{Kramm, U.}, \bibinfo{author}{Nettelmann, N.},
  \bibinfo{author}{Redmer, R.}, \bibinfo{author}{Stevenson, D.J.},
  \bibinfo{year}{2011}.
\newblock \bibinfo{title}{{Astrophysics On the degeneracy of the tidal Love
  number $k_2$ in multi-layer planetary models : application to Saturn and GJ
  436b}}.
\newblock \bibinfo{journal}{Astron. Astrophys.} \bibinfo{volume}{18},
  \bibinfo{pages}{1--7}.
\newblock \DOIprefix\doi{10.1051/0004-6361/201015803},
  \href{http://arxiv.org/abs/1101.0997}{\tt arXiv:1101.0997}.
\bibitem[{Lainey et~al.(2016)Lainey, Jacobson, Tajeddine, Cooper, Robert,
  Tobie, Guillot and Mathis}]{lainey2016}
\bibinfo{author}{Lainey, V.}, \bibinfo{author}{Jacobson, R.A.},
  \bibinfo{author}{Tajeddine, R.}, \bibinfo{author}{Cooper, N.J.},
  \bibinfo{author}{Robert, V.}, \bibinfo{author}{Tobie, G.},
  \bibinfo{author}{Guillot, T.}, \bibinfo{author}{Mathis, S.},
  \bibinfo{year}{2016}.
\newblock \bibinfo{title}{{New constraints on Saturn's interior from Cassini
  astrometric data}} \URLprefix \url{http://arxiv.org/abs/1510.05870},
  \href{http://arxiv.org/abs/1510.05870}{\tt arXiv:1510.05870}.
\bibitem[{Munk and MacDonald(2009)}]{munk2009}
\bibinfo{author}{Munk, W.H.}, \bibinfo{author}{MacDonald, G.J.F.},
  \bibinfo{year}{2009}.
\newblock \bibinfo{title}{{The Rotation of the Earth: A Geophysical
  Discussion}}.
\newblock Cambridge Monographs on Mechanics, \bibinfo{publisher}{Cambridge
  University Press}.
\newblock \URLprefix \url{https://books.google.com/books?id=klDqPAAACAAJ}.
\bibitem[{Nettelmann et~al.(2013)Nettelmann, P{\"{u}}stow and
  Redmer}]{Nettelmann2013}
\bibinfo{author}{Nettelmann, N.}, \bibinfo{author}{P{\"{u}}stow, R.},
  \bibinfo{author}{Redmer, R.}, \bibinfo{year}{2013}.
\newblock \bibinfo{title}{{Saturn layered structure and homogeneous evolution
  models with different EOSs}}.
\newblock \bibinfo{journal}{Icarus} \bibinfo{volume}{225},
  \bibinfo{pages}{548--557}.
\newblock \URLprefix
  \url{http://linkinghub.elsevier.com/retrieve/pii/S0019103513001784},
  \DOIprefix\doi{10.1016/j.icarus.2013.04.018},
  \href{http://arxiv.org/abs/1304.4707}{\tt arXiv:1304.4707}.
\bibitem[{Tassoul(2015)}]{tassoul2015}
\bibinfo{author}{Tassoul, J.L.}, \bibinfo{year}{2015}.
\newblock \bibinfo{title}{{Theory of Rotating Stars. (PSA-1)}}.
\newblock Princeton Series in Astrophysics, \bibinfo{publisher}{Princeton
  University Press}.
\newblock \URLprefix \url{https://books.google.com/books?id=nnJ9BgAAQBAJ}.
\bibitem[{Wisdom(1996)}]{wisdom1996}
\bibinfo{author}{Wisdom, J.}, \bibinfo{year}{1996}.
\newblock \bibinfo{title}{{Non-perturbative Hydrostatic Equilibrium}}
  \URLprefix \url{http://web.mit.edu/wisdom/www/interior.pdf}.
\bibitem[{Wisdom and Hubbard(2016)}]{wisdom2016}
\bibinfo{author}{Wisdom, J.}, \bibinfo{author}{Hubbard, W.B.},
  \bibinfo{year}{2016}.
\newblock \bibinfo{title}{{Differential rotation in Jupiter: A comparison of
  methods}}.
\newblock \bibinfo{journal}{Icarus} \bibinfo{volume}{267},
  \bibinfo{pages}{315--322}.
\newblock \URLprefix \url{http://dx.doi.org/10.1016/j.icarus.2015.12.030},
  \DOIprefix\doi{10.1016/j.icarus.2015.12.030}.
\bibitem[{Zharkov and Trubitsyn(1978)}]{zharkov1978}
\bibinfo{author}{Zharkov, V.N.}, \bibinfo{author}{Trubitsyn, V.P.},
  \bibinfo{year}{1978}.
\newblock \bibinfo{title}{{The physics of planetary interiors}}.
\newblock \bibinfo{publisher}{Parchart}, \bibinfo{address}{Tucson, AZ}.

\end{thebibliography}
\bibliographystyle{model2-names.bst}\biboptions{authoryear}

\clearpage


\begin{deluxetable}{cccc}
\tablewidth{0pc}
\tablecaption{Comparing two-layer models}
\tablehead{
    \colhead{$a_1/a$}   &  \colhead{$\rho_0/\rho_1$}  &  
    \colhead{$k_2$~CMS}  &  \colhead{$k_2$~Clairaut}  }
\startdata
0.1       &  0.5             &  1.496283   &  1.496286        \\
0.3       &  0.5             &  1.411183   &  1.411185        \\
0.5       &  0.1             &  0.465714   &  0.465716        \\
0.5       &  0.3             &  0.947967   &  0.947969        \\
0.5       &  0.5             &  1.205309   &  1.205311        \\
0.5       &  0.7             &  1.360183   &  1.360186        \\
0.5       &  0.9             &  1.461667   &  1.461669        \\
0.7       &  0.5             &  1.057405   &  1.057407        \\
0.9       &  0.5             &  1.217192   &  1.217194        \\
\enddata
\tablecomments{ Calculated $k_2$ for a two layer model with $q_{\rm tid}=10^{-6}$,
$q_{\rm rot}=0$ and Tethy's $R/a$, for chosen values of ratio of radii and densities
of the two layers. Results closely match the approximation using Clairaut theory in
\citet{folonier2015}, Eqn. 41.}
\label{tab:folonier_table}
\end{deluxetable}

\begin{deluxetable}{rrrr}
\tablecolumns{3}
\tablewidth{0pc}
\tablecaption{Saturn Model Parameters}
\tablehead{
\colhead{} & Cassini  & Voyager & \colhead{} }
\startdata
$GM$ & $3.7931208 \times 10^{7}$ \tablenotemark{a} & \nodata & ${\rm (km^3/s^2) }$
\\
$a$ & $6.0330 \times 10^{4}$       \tablenotemark{a} & \nodata & ${\rm (km) }$  \\ 
$J_2  \times  10^6$  &  $16290.71$ \tablenotemark{a} &  \nodata  &       \\
$J_4  \times  10^6$  &  $-935.83$  \tablenotemark{a} &  \nodata  &       \\
$J_6  \times  10^6$  &  $86.14$    \tablenotemark{a} &  \nodata  &       \\
$q_{\rm rot} $  & $0.1516163$ \tablenotemark{b} & $0.1553029$ \tablenotemark{c} &\\
$r_{\rm core} / a$ & $0.2$ & \nodata &  \\ 
$m_{\rm core} / M$  &  $0.133146$ & $0.140478$ &\\
\tableline
\tableline
  & Tethys  & Dione & \\
\tableline
$q_{\rm tid} $  & $-2.791103 \times 10^{-8} $ \tablenotemark{d}  
& $-2.364582 \times 10^{-8}$ \tablenotemark{d} & \\
$R/a $  & $4.8892$ \tablenotemark{d} & $6.2620$ \tablenotemark{d} &  
\enddata
\tablerefs{ a. \citet{Jacobson2006}, b. \citet{giampieri2006}, c. \citet{desch1981},
d.  \citet{archinal2011}}
\tablecomments{ Identical parameters for Saturn are used with the exception of
    $q_{\rm rot}$, for which the rotation rate from both \textit{Cassini} and
    \textit{Voyager} are considered. A constant core density is fitted to match $J_2$,
    $J_4$, and $J_6$ for a converged figure. }
\label{tab:saturn_params}
\end{deluxetable}

\begin{deluxetable}{ccrcr}
\tablecolumns{3}
\tablewidth{0pc}
\tablecaption{Calculated Saturn tidal responses}
\tablehead{
\colhead{model} &  & gravitational moment  & & normalized moment}
\startdata
\colhead{Cassini} &  $J_2$     &  $1.62907100025\times10^{-2}   $  &  $J_2/q_{\rm  rot}$  &  $0.10744694879478             $  \\
 \colhead{no~tide}   &  $J_4$     &  $-9.2027941201\times10^{-4}   $  &  $J_4/q_{\rm  rot}$  &  $-0.606979160784\times10^{-2} $  \\
 \colhead{$N=128$}  &  $J_6$     &  $8.014294995\times10^{-5}     $  &  $J_6/q_{\rm  rot}$  &  $0.5285905549\times10^{-3}    $  \\
\tableline
\colhead{non-rotating}  & $C_{22}$  &  $8.5288\times10^{-10}         $  &  $k_2         $      &  $0.36669                      $ \\
\colhead{Tethys}        & $J_2$     &  $1.70576\times10^{-9}         $  &  $J_2/q_{\rm  rot}$  &  \colhead{\nodata}               \\
\colhead{$N=128$}       & $J_4$     &  $-1.351\times10^{-11}         $  &  $J_4/q_{\rm  rot}$  &  \colhead{\nodata}               \\
                        & $J_6$     &  $2.2\times10^{-13}            $  &  $J_6/q_{\rm  rot}$  &  \colhead{\nodata}               \\
\tableline
\colhead{Cassini}  & $C_{22}$  &  $9.6070\times10^{-10}         $  &  $k_2         $      &  $0.41304                      $ \\
\colhead{Tethys}   & $J_2$     &  $1.629071017501\times10^{-2}  $  &  $J_2/q_{\rm  rot}$  &  $0.1074469499328              $ \\
\colhead{$N=128$}  & $J_4$     &  $-9.2027943932\times10^{-4}   $  &  $J_4/q_{\rm  rot}$  &  $-0.60697917880\times10^{-2}  $ \\
                   & $J_6$     &  $8.01429541\times10^{-5}      $  &  $J_6/q_{\rm  rot}$  &  $0.5285905822\times10^{-3}    $ \\
\tableline
\colhead{Voyager}  & $C_{22}$  &  $9.4136\times10^{-10}         $  &  $k_2         $      &  $0.40473                      $ \\
\colhead{Tethys}   & $J_2$     &  $1.629071048760\times10^{-2}  $  &  $J_2/q_{\rm  rot}$  &  $0.1048963407747              $ \\
\colhead{$N=128$}  & $J_4$     &  $-9.3570887868\times10^{-4}   $  &  $J_4/q_{\rm  rot}$  &  $-0.60250556585\times10^{-2}  $ \\
                   & $J_6$     &  $8.30176108\times10^{-5}      $  &  $J_6/q_{\rm  rot}$  &  $0.534552720\times10^{-3}     $ \\
\tableline
\colhead{Cassini}  & $C_{22}$  &  $8.1325\times10^{-10}         $  &  $k_2         $      &  $0.41272                      $ \\
\colhead{Dione}    & $J_2$     &  $1.629071019035\times10^{-2}  $  &  $J_2/q_{\rm  rot}$  &  $0.1074469500340              $ \\
\colhead{$N=128$}  & $J_4$     &  $-9.2027943688\times10^{-4}   $  &  $J_4/q_{\rm  rot}$  &  $-0.60697917719\times10^{-2}  $ \\
                   & $J_6$     &  $8.01429534\times10^{-5}      $  &  $J_6/q_{\rm  rot}$  &  $0.528590578\times10^{-3}     $ \\
\tableline
\colhead{Cassini}   &  $C_{22}$  &  $9.6219\times10^{-10}         $  &  $k_2         $      &  $0.41368                      $  \\
\colhead{Tethys}    &  $J_2$     &  $1.629071019560\times10^{-2}  $  &  $J_2/q_{\rm  rot}$  &  $0.1074469500686              $  \\
\colhead{$N=4$}     &  $J_4$     &  $-9.3583002600\times10^{-4}   $  &  $J_4/q_{\rm  rot}$  &  $-0.61723571821\times10^{-2}  $  \\
                    &  $J_6$     &  $8.61400043\times10^{-5}      $  &  $J_6/q_{\rm  rot}$  &  $0.568144705\times10^{-3}     $  \\
\enddata
\label{tab:saturn_results}
\end{deluxetable}

\end{document}